\def\year{2023}\relax
\documentclass[letterpaper]{article} % DO NOT CHANGE THIS

\usepackage{xspace}

\usepackage{aaai22}  % DO NOT CHANGE THIS
\usepackage{times}  % DO NOT CHANGE THIS
\usepackage{helvet}  % DO NOT CHANGE THIS
\usepackage{courier}  % DO NOT CHANGE THIS
\usepackage[hyphens]{url}  % DO NOT CHANGE THIS
\usepackage{graphicx} % DO NOT CHANGE THIS
\usepackage{natbib}  % DO NOT CHANGE THIS AND DO NOT ADD ANY OPTIONS TO IT
\usepackage{caption} % DO NOT CHANGE THIS AND DO NOT ADD ANY OPTIONS TO IT
\DeclareCaptionStyle{ruled}{labelfont=normalfont,labelsep=colon,strut=off} % DO NOT CHANGE THIS
\usepackage{algorithm}
\usepackage{algorithmic}

\usepackage{newfloat}
\usepackage{listings}
\floatstyle{ruled}
\newfloat{listing}{tb}{lst}{}
\floatname{listing}{Listing}
\title{Party Prediction for Twitter}
\author{
Kellin Pelrine\textsuperscript{1,2}, Anne Imouza\textsuperscript{1,2}, Zachary Yang\textsuperscript{1,2}, Jacob-Junqi Tian\textsuperscript{1,2}, Sacha L\'evy\textsuperscript{1,2}, Gabrielle Desrosiers-Brisebois\textsuperscript{2,3}, Aarash Feizi\textsuperscript{1,2}, C\'ecile Amadoro\textsuperscript{3}, Andr\'e Blais\textsuperscript{3}, Jean-Fran\c{c}ois Godbout\textsuperscript{3}, Reihaneh Rabbany\textsuperscript{1,2}
    }
\affiliations{
    \textsuperscript{\rm 1}McGill University \\
    \textsuperscript{\rm 2}Mila Institute\\
    \textsuperscript{\rm 3}Universit\'e de Montr\'eal\\
    }

\AtBeginDocument{%
  }

\usepackage{amsmath}
\usepackage{amsfonts}
\usepackage{amssymb}
\usepackage{graphicx}
\usepackage{color}
\usepackage{xcolor,colortbl}
\usepackage{url}
\usepackage{paralist} % also compact lists
\usepackage{listings} % to get sql etc nicely formatted
\usepackage{bbding} % for the \CheckmarkBold symbol
\usepackage[draft]{changes} % allow for orange-suggestions
\usepackage{natbib} 
\usepackage{lipsum}
\usepackage{graphicx}
\usepackage{tabularx}
\usepackage[normalem]{ulem}
\useunder{\uline}{\ul}{}
\usepackage{booktabs}
\usepackage{pifont}
\usepackage{array}
\newcolumntype{H}{>{\setbox0=\hbox\bgroup}c<{\egroup}@{}}

\newcommand*\X{\ding{51}}
\newcommand*\YES{\ding{51}}
\newcommand*\NO{ }

\usepackage{tikz}
\usepackage{diagbox}

\usepackage{multirow}

\newcommand{\hide}[1]{}
 \definecolor{brickred}{rgb}{0.8, 0.25, 0.33}

\newcommand{\bit}{\begin{compactitem}}
\newcommand{\eit}{\end{compactitem}}
\newcommand{\ben}{\begin{compactenum}}
\newcommand{\een}{\end{compactenum}}

\newcommand{\Public}{US Public\xspace}
\newcommand{\Politicians}{US Politicians\xspace}
\newcommand{\Canadian}{CA Public\xspace}

\usepackage{xcolor}
\usepackage{pgf}
\usepackage{collcell}
\usepackage{enumitem}
\usepackage{wrapfig}
\usepackage{xstring}

\newcommand*{\MinNumber}{37}%
\newcommand*{\MaxNumber}{1}%

\newcommand{\ApplyGradient}[1]{
        \IfDecimal{#1}{%
            \pgfmathsetmacro{\PercentColor}{min(100,max(0,100.0*(#1-\MinNumber)/(\MaxNumber-\MinNumber)))}%
            \textcolor{blue!\PercentColor!orange}{#1}
        }{#1}% if it's not a number, just print it as is
}

\newcolumntype{R}{>{\collectcell\ApplyGradient}{r}<{\endcollectcell}}

\date{}
\newif\ifblind

\begin{document}
\newcommand{\method}{\textsc{our method}\xspace}

\maketitle

\begin{abstract}
A large number of studies on social media compare the behaviour of users from different political parties. As a basic step, they employ a predictive model for inferring their political affiliation. The accuracy of this model can change the conclusions of a downstream analysis significantly, yet the choice between different models seems to be made arbitrarily. In this paper, we provide a comprehensive survey and an empirical comparison of the current party prediction  practices and propose several new approaches which are competitive with or outperform state-of-the-art methods, yet require less computational resources. Party prediction models rely on the content generated by the users (e.g., tweet texts), the relations they have (e.g., who they follow), or their activities and interactions (e.g., which tweets they like). We examine all of these and compare their signal strength for the party prediction task. This paper lets the practitioner select from a wide range of data types that all give strong performance. Finally, we conduct extensive experiments on different aspects of these methods, such as data collection speed and transfer capabilities, which can provide further insights for both applied and methodological research.

\end{abstract}

\section{Introduction}
\label{sec:intro}
% use about 1 page, start with a rough draft and update when other sections are all written
% One paragraph, motivation intro for general CS audience
% Second paragraph narrow down on the problem, briefly summarize the gaps in the state of the art and current challenges 
% third paragraph introduce the solution presented by our paper
% fourth paragraph give an illustration of how the proposed method preforms 
% enumerate the contributions clearly with bullet points
% declare the code location, provide a github link
% by the end of intro, the reader should get an overall idea of the problem, method, its significance, and performance

Inferring the political orientation of social media users is a common first step in a wide range of studies. Multiple works on understanding the impact and the spread of misinformation consider the political ideology of users \cite{osmundsen_bor_vahlstrup_bechmann_petersen_2021,lawson2022pandemics} in different contexts to determine, for example, how political ideology can predict amplifying fake news originating from trolls \cite{badawy2019falls}. % It is fundamental to polling and predicting election outcomes. 
Another critical application of party prediction is for measuring political conflicts in online societies \cite{green2020elusive, conover2011predicting}, and for devising strategies to mitigate polarization's harms \cite{voelkelmegastudy,ruggeri2021general,bail2018exposure,warner2017test}. Indeed, numerous studies have confirmed that partisan animosity and polarization has increased in several countries, such as the United States \citep{tucker2018social}, Canada \citep{johnston2019affective}, the United Kingdom \citep{hobolt2021divided}, and France \citep{boxell2022cross}, with researchers identifying social media as one of the root cause of this phenomenon \citep{kubin2021role, banks2021polarizedfeeds, kim2019incivility, heiss2019populist, pnas2022tornberg}. 
However, drawing robust conclusions about the impacts of political ideology and how it changes requires that researchers correctly identify the partisan preferences of social media users. While this is a crucial and extensively studied topic, there is currently no widely accepted or definitive method for predicting an individual's political party affiliation based on their online activity.

In this paper, we discuss how existing methods for predicting political party affiliation of social media users are generally based on several different types of signals, including content (e.g., text of tweets: \citealt{conover2011predicting,rheault2021efficient,colleoni2014echo}. Or linked media: \citealt{luceri2019red,pnas2022tornberg}), relationships (e.g., followership: \citealt{barbera2015birds,gu2016ideology}), and interactions (e.g., retweeting: \citealt{xiao2020timme,badawy2018analyzing}). Most of the previous methods use a combination of these features, as summarized in Table \ref{tab:literature_survey} and continued with Table 10 in the supplementary material; they are also normally evaluated on original datasets associated with broader studies. The collection process and resulting data difficulty vary widely, with different types of users (e.g., politicians vs. general public), diverse filters, and often entirely different time periods and topics studied. This means that assessing the performance of these models in the literature is challenging, as there is usually no empirical comparison done on the performance of the utilized methods versus other existing approaches or baselines. Although this can be expected from larger systems, sub-optimal or uncertain party prediction performance can have a big impact on the downstream results and overall conclusions.% Thus, there are few comparisons within each paper, and evaluations of the different approaches are generally challenging due to the diverse data sources.

% To address this problem, we first survey the literature, paying particular attention to not only the method and its aggregate performance numbers, but also the characteristics of the data used. We summarize the key findings in Tables \ref{tab:literature_survey} and \ref{tab:literature_survey_appendix}, and elaborate in detail in Section 2. 

%###ALTERNATIVE MERGED PARAGRAPHS
We address these shortcomings by conducting a comprehensive survey of existing party prediction methods, followed by a thorough evaluation through an extensive array of experiments. The goal of this comparison is to find the most effective and practical methods, so we go beyond mere \textit{accuracy} and consider also \textit{coverage} and \textit{cost}. Coverage measures how many users we can classify with each method, whereas cost focuses on how much resources are required to acquire the relevant data for this particular method. For example, requiring only tweet stream vs. full followership network: the latter is commonly used, but it requires more extensive data collection while not performing better than less data intensive alternatives, as we discuss in the experiments.

In order to conduct this analysis, we collected a rich dataset of approximately 14,000 Twitter users who discuss American politics during the time period directly before and after the 2020 US election. This dataset includes 8 distinct signals which enables a unified comparison to test the performance of different methods. We also propose our own approaches based on label propagation, GCN (Graph Convolutional Networks), GAT (Graph Attention Networks), HAN (Heterogeneous Graph Attention Networks), and RoBERTa (a transformer-based language model), and show that these methods deliver strong performance and can do so from a wide variety of data types.
%######
%
%WOULD NEED TO DELETE BELOW IF USE THE PARAGRAPHS ABOVE
% To address this problem, this paper presents the first comprehensive comparative analysis to evaluate the performance of these models. For this, we collect a rich dataset of approximately 14,000 Twitter users who discuss American politics during the time period directly before and after the 2020 US election. This dataset includes distinct signals which enables a unified comparison to test the performance of different methods. We also propose our own approaches based on XXX, and confirm that our methods deliver strong performance and can do so from a wide variety of data types. %In this way we confirm quantitatively the major impact of different datasets and how challenging the task of comparing existing models is. 
In summary, the main contributions are:
% Moreover, our study provides the missing thorough comparison necessary to evaluate the performance of  these models for the first time. 
%
% We also do extensive ablation and other experiments validating different components of our approaches. We confirm that our methods deliver strong performance and can do so from a wide variety of data types. Our experiments can also help future researchers better understand all the different factors involved in this task, such as which data types are most informative and applicable, how to structure interaction data, and differences between the public and politicians. We hope that in this way our experiments will lead to improved methods and measures of online partisan conflict. 
%

\begin{itemize}[leftmargin=10pt,topsep=0pt]
\item To provide a thorough comparison of different methods for party prediction, including the four current state-of-the-art methods as well as thirty-four newly proposed methods which rely on different combinations of text, activity and relation signals.
\item To show that our proposed label propagation on retweet activities has a strong accuracy, coverage and speed, while its data acquisition is also the most efficient. The coverage of this method can be improved by combining more signals and/or using more complex methods.
\item To open up new options and data types for applied practitioners and new insights for future methodological research. For example, in our experiments on Canadian data the followership information was not available, hence the party prediction was performed using methods that rely on retweets and mentions. Having more options like these will be increasingly important as Twitter data access becomes more uncertain and some data types become inaccessible or more costly to collect.
\end{itemize}
Ultimately, our experiments should help researchers better understand the myriad of factors involved in this task, such as which data types are most informative and applicable, how to structure interaction data, and differences between the accounts of general users and those of politicians.
\noindent Our code and supplementary materials are available at https://github.com/anonymouspartyprediction/partypred/. %\href{https://anonymous.4open.science/r/Party\_Prediction-395D/}{github}.

\begin{table*}[th!]
\caption{Survey methods to predict party affiliation. Here we report for each paper:  Accuracy, whether they used Media outlets, Network Activities (retweets and mentions), Network Relation (followership), and Content  (words and  hashtags); as well as if they consider  Public, Elite or both, their test size in terms of number of users on which the ideology is inferred, if their code is available, and finally the level of difficulty based on the type and size of data (elaborated in the text). %, as explained more in the text.  
}\vspace{-9pt}
\label{tab:literature_survey}
\resizebox{1\linewidth}{!}
{%
\begin{tabular}{l|c|cccc|lrcc}
 
\textbf{Methods} &
%   \rotb{Accuracy (1)} &
%   \rotb{Media outlets (2)} &
%   \rotb{\begin{tabular}[c]{@{}c@{}}Network Activities\\   (retweets and mentions) (3)\end{tabular}} &
%   \rotb{\begin{tabular}[c]{@{}c@{}}Network Relation\\ (fellowship) (4)\end{tabular}} &
%   \rotb{\begin{tabular}[c]{@{}c@{}}Content \\ (words and  hashtags) (5) \end{tabular}} &
%   \rotb{Public vs. Elite} &
%   \rotb{Test size} &
%   \rotb{Code available?} &
%   \rotb{Level of Difficulty} \\ 
  \textbf{Accuracy} &
  \textbf{Media} &
  \textbf{\begin{tabular}[c]{@{}c@{}}Activity \end{tabular}} &
  \textbf{\begin{tabular}[c]{@{}c@{}}Relation \end{tabular}} &
  \textbf{\begin{tabular}[c]{@{}c@{}}Content  \end{tabular}} &
  \textbf{Type} &
  \textbf{Size} & % (users)} &
  \textbf{Code} &
  \textbf{Difficulty} \\ 
  %\multicolumn{1}{c}{\textbf{\begin{tabular}[c]{@{}c@{}}Dataset Difficulty: \\ (how many users? any filter on users?\\  type (politicians or their followers), \\ type \& amount of activity,\\  keywords they have use?)\end{tabular}}} &
  \hline
\citet{conover2011predicting}%r et al. (2011) 
&
  95   &
   &
  \X &
   &
  \X &
  Public &
  1,000  &
  \NO &
 % \begin{tabular}[c]{@{}l@{}}* Users well-connected \\ * At least one political hashtag.\end{tabular} &
  Medium \\
  \hline
\textbf{\citet{barbera2015birds}} %rá et al. (2015) 
&
  78   &
   &
   &
  \X &
   &
  Both &
  42,008   &
  %\begin{tabular}[c]{@{}l@{}}* Users with more than 100 tweets.\\ * 25 or more followers.\\ * Follow 100 or more. \\ * English.\\ * Mention collection keywords. \\ *(Non) Political hashtags.\end{tabular} &
  \multicolumn{1}{c}{\YES} & %\href{https://github.com/pablobarbera/echo\_chambers}{\YES}
  \textbf{Hard} \\
  \hline
   
\citet{rheault2021efficient}
% and Musulan (2021)
&
  91   &
  \X &
  \X &
   &
  \X &
  Politicians &
  505  &
  %\begin{tabular}[c]{@{}l@{}}* Hashtags referring to the \\ 2019 Canadian election\\ * True party affiliation\end{tabular} &
   \multicolumn{1}{c}{\YES} & % \href{https://dataverse.harvard.edu/dataset.xhtml?persistentId=doi:10.7910/DVN/P6SZ2G}{\YES}
   Easy \\
   \hline
    
\citet{luceri2019red}
% i et al. (2019)
&
  89   &
  \X &
  \X &
   &
   &
  Public &
  38,000  &
  \NO &
  %\begin{tabular}[c]{@{}l@{}}*Other 1 million accounts labelled\\ with the retweet network of the users\end{tabular} &
  \begin{tabular}[c]{@{}c@{}} Medium \end{tabular} \\
  \hline
   
\citet{colleoni2014echo}
% ni et al. (2014) 
&
  79   &
   &
   &
  \X &
  \X &
  Public &
  10,551  &
  \NO &
  %\begin{tabular}[c]{@{}l@{}}* Political tweets\\ * Follow Democrat or Republican\\  accounts.\end{tabular} &
  Medium \\
  \hline
   
\citet{pennacchiotti2011machine}
% acchiotti and Popescu (2011)
&
  89   &
   &
  \X &
  \X &
  \X &
  Public &
  10,338 &
  \NO &
  %* Classified themselves. &
  Medium \\
  \hline
   
% Stefanov et al. (2020) 
\citet{stefanov2020predicting}
&
  83   &
  \X &
  \X &
   &
  \X&
  Public &
  806  &
  \NO &
  %\begin{tabular}[c]{@{}l@{}}* Discuss one of 8 eight political topics.\\ * Located in the USA.\\ * URLs from media with known bias.\end{tabular} &
  Medium \\
  \hline
   
% Gu et al. (2016)
\citet{gu2016ideology}
&
  66   &
   &
  \X &
  \X &
   &
  Both &
  1,200  &
  \NO &
  Medium 
 % \begin{tabular}[c]{@{}l@{}}* Users connected to a set of politicians.\\ * Followed by at least 20 politicians \\ * Follow only 3 to 5  politicians.\end{tabular} & N0
   \\
   \hline
    
% Badawy et al. (2018) 
\citet{badawy2018analyzing}&
  91   &
  \X &
  \X &
   &
   &
  Public &
  29,000 &
  \NO &
  %\begin{tabular}[c]{@{}l@{}}* Retweet messages from left or right media.\\ * Keywords related to U.S. Presidential election.\\ *Keywords related to D.Trump and H. Clinton.\\ *Keywords for other parties added.\end{tabular} &
  Medium \\
  \hline
   
% Xiao et al. (2020)
\textbf{\citet{xiao2020timme}}&
  \textbf{96}   &
   &
  \X &
  \X &
   &
  Both &
  20,811  &
 % \begin{tabular}[c]{@{}l@{}}* Accounts of politicians and candidates. \\ * Crawling of every tweet of each user (public).\\ * Keen on political affairs and moderate interests.\end{tabular} &
  \multicolumn{1}{c}{\YES} & % \href{https://github.com/PatriciaXiao/TIMME}{\YES}
  \textbf{Hard}\\
  \hline
   
  \textbf{\citet{preoctiuc2017beyond}} &
  \textbf{97}   &
   &
   &
   \X &
   \X&
   Public &
   13,651  &
   \multicolumn{1}{c}{\YES} & % \href{http://www.preotiuc.ro/resources.html}{\YES}
   Medium \\
   \hline
   \citet{rodriguez2022urjc}
% and Musulan (2021)
&
  82   &
   &
   &
   &
  \X &%1029600
  Politicians  &
  12,600  &
  %\begin{tabular}[c]{@{}l@{}}* Hashtags referring to the \\ 2019 Canadian election\\ * True party affiliation\end{tabular} &
   &
   Easy \\
      \hline
% \cit{rodriguez2022urjc}
% % and Musulan (2021)
% &
%   84   &
%   &
%   &
%   &
%   \X &
%   Politicians (left) &
%   6,840  &
%   %\begin{tabular}[c]{@{}l@{}}* Hashtags referring to the \\ 2019 Canadian election\\ * True party affiliation\end{tabular} &
%   &
%   Easy \\
%       \hline
   
% \cit{rodriguez2022urjc}
% % and Musulan (2021)
% &
%   79   &
%   &
%   &
%   &
%   \X &
%   Politicians (right) &
%   5,760  &
%   %\begin{tabular}[c]{@{}l@{}}* Hashtags referring to the \\ 2019 Canadian election\\ * True party affiliation\end{tabular} &
%   &
%   Easy \\
%     \hline
   
\citet{mou2021align}
% and Musulan (2021)
&
  80   &
   &
   &
  \X &
  \X &
  Politicians &
  735  &
  %\begin{tabular}[c]{@{}l@{}}* Hashtags referring to the \\ 2019 Canadian election\\ * True party affiliation\end{tabular} &
  \multicolumn{1}{c}{\YES} % \href{https://github.com/xymou/Align-Voting-Behavior-with-Public-Statements/tree/main/code/data}{\YES}
   &
   Easy \\
\end{tabular}}\vspace{-5pt}
\end{table*}

\section{Background and Related Work}
\label{sec:background}

Party prediction is a common approach used to classify social media users according to their partisan affiliations. In the United States, for example, users can either be supporters of the Democratic or the Republican parties; in some cases, users can also be classified according to their ideology, as liberals or conservatives. Although extensively used \cite{lawson2022pandemics,badawy2019falls,green2020elusive, conover2011predicting,voelkelmegastudy,ruggeri2021general,bail2018exposure,warner2017test}, there is no widely accepted consensus among scholars on the right approach to conduct this classification task. % \citep{cohen2013classifying}. 
Most existing methods have been applied to Twitter data due to both the ease of access for researchers, as well as its mainstream use for political discourse. In this study, we also use Twitter data to compare methods for party prediction. However, the conclusions could transfer to similar media platforms.

There is currently a broad range of estimation techniques applied for party prediction on Twitter. Some party prediction approaches use media outlets shared by the users to infer partisan affiliation \citep{rheault2021efficient, luceri2019red, stefanov2020predicting, badawy2018analyzing}. Others rely on the structure of networks to infer partisan leanings \citep{Barber2015TweetingFL, colleoni2014echo, pennacchiotti2011machine, gu2016ideology, xiao2020timme, havey2020partisan, wojcieszak2022most}, while some rely instead on other types of network activities, such as retweets and mentions \citep{conover2011predicting, rheault2021efficient, luceri2019red, pennacchiotti2011machine, gu2016ideology, badawy2018analyzing, xiao2020timme, jiang2021social}. Finally, several approaches focus on the content of messages, by either looking at the text or the hashtags used to represent a partisan side, like \#Democrats or \#Republican \citep{conover2011predicting, rheault2021efficient, colleoni2014echo, pennacchiotti2011machine, rodriguez2022urjc, mou2021align, fagni2022fine}. Many works combine more than one feature to make their predictions \citep{Makazhanov&Rafiei2013, boutet2012s, wang2017polarized, mou2021align}, while a few rely exclusively on a single feature in their party prediction model (e.g., \citealt{Barber2015TweetingFL}). 
 
We summarize some of these key methods proposed in the literature for party prediction in Table \ref{tab:literature_survey}, based on what signals they use to infer the user's ideology as well as the distribution of the users they classified. For the former, we consider whether they use (column 2) media outlets (shared by the user in their messages) (3) network activities (who the user interacts with through retweets and mentions) (4) network relations (who the user follows) and (5) content (hashtags and keywords in their messages). We also look into the difficulty of users classified, by reporting the (6) type of user (politician or public), (7) number of users and (9) our estimated difficulty as explained below.
% But how accurate are these different classification approaches? Since the goal of this paper is to develop a standard measure of party affiliation from online activities, comparing our own approach to other competing techniques is a necessary step to validate its performance. For the purpose of this study, we reviewed 20 papers and identified eleven approaches that offer a unique method to classify users according to their partisan affiliation. We have selected these approaches because they represent a broad range of estimation techniques and because they offer strong performances to predict the partisan affiliation of social network users, with the most promising of these methods tested against different gold standards to establish their accuracy. 
% As a quick overview,
Methods selected for inclusion in Table~\ref{tab:literature_survey} represent a broad range of estimation techniques, and generally report high levels of prediction accuracy. Even so, these levels of accuracy fall between 66\% and 97\%, meaning that there is a wide range of classification error, depending on which approach is used. 
In the supplementary material we have a similar table for additional notable related works which self-report lower accuracy levels. %We selected the models for inclusion in Table~\ref{tab:literature_survey} based on their higher level of prediction accuracy (as reported by the authors). 
% Even so, 
% We explain these tables below. 

%Each classification technique is based on Twitter data and uses different features to assess the partisan affiliation of users. We rate each of these approach according to their levels of difficulty, meaning that we evaluate the method applied in each of the papers according to a series of criteria and then assign them a level of difficulty ``easy'', ``medium'', or ``hard''.

% To compare the methods, we report (1) the accuracy as reported by the authors, %which indicated how well the party predictions perform (). 
% the features  used in the classification: (2) media outlets (3) network activities (4) network relations or (5) content. 

For comparing party prediction methods, it is important to consider if the classification task was performed within a subsample of ``politicians'' or the general ``public''; that is, accounts from well-known politicians and parties (see for example \citealt{rheault2021efficient, Barber2015TweetingFL, gu2016ideology, xiao2020timme}), or from a broader set of Twitter users, which can be harder to classify. Even when classification extends to the public, the way these users are sampled could impact the difficulty of the classification task (e.g., sampling only hyper-partisan or super active users which are easier to classify). Therefore, to aid comparison, we have provided an overall level of difficulty for each method's test data by investigating their sampling approach. However, this level of difficulty is approximate and cannot give a precise, fine-grained comparison. This motivates our empirical analysis where we compare methods on the same dataset. %Note that all methods are not easily reproducible. Therefore, we have determined an overall level of difficulty for each method by investigating their sampling approach to provide a more comprehensive comparison. 
%\footnote{%To determine this overall level of difficulty of the classification tasks, we proceeded in four steps. First, we examined whether the models were tested on a general sample of Twitter users or if they focused on politician accounts instead. Second, we looked at the number of filters applied to the data before the classification task. 

Almost all of the previous methods are tested on a sample of users who discuss political topics (i.e., datasets selected from policy-related keywords or hashtags or tweets of known politicians). However some apply additional filters (e.g., selection on polarizing keywords, user activity levels, or user location) as a factor that could potentially change the difficulty level. %Third, we examined the data collection approach by considering the number of features included in the classification task; methods that integrate several features are assumed to be more complex in their implementation. Fourth, we considered how each approach was tested, such as on the type of data and on the sample size of users and tweets. Here, the larger the sample or test size, the more complex the classification task. From this, we have labelled all approaches as either ``easy'', ``medium'', or ``hard''.} 
% We also added a column (7) related to the test size (i.e., the number of users an approach was evaluated on) to validate the accuracy of the method which varies from around 500 \citep{rheault2021efficient} to approximately 40,000 \citep{Barber2015TweetingFL, luceri2019red}. Finally, the last two columns indicate whether the code used to make those predictions is publicly available (8), and our own evaluation of the difficulty level of the prediction task and data which the method was tested on (9).
In our analysis, models tested exclusively on politicians are assigned an \textit{``easy''} difficulty level since these types of users are known to be easier to classify \citep{cohen2013classifying, rodriguez2022urjc}.  
% Only one paper falls into this category \citep{rheault2021efficient}. 
A \textit{``medium''} level of difficulty implies that the method applies to a broader set of users from the general public, and contains more complex features \citep{conover2011predicting, luceri2019red, colleoni2014echo, pennacchiotti2011machine, stefanov2020predicting, badawy2018analyzing,gu2016ideology}. An example here would be \citet{stefanov2020predicting} who classify general public users based on the ideological leanings of the news articles they share on Twitter. For the classification task to be \textit{``hard''}, the test size must also be large and validation must be done on both the public and an elite group of politicians. %This category also implies that the method used is sophisticated, either because of the test sample size or the features used. 
Two papers fall into this category \citep{Barber2015TweetingFL, xiao2020timme}, which we use later on in the experiments as baselines. Notably, the method proposed in \citet{Barber2015TweetingFL}, based on followership, is one of the most commonly used measure to estimate the levels of partisanship of users at the individual level \citep{brady2017emotion, jost2018social, pennycook2021shifting}. The overall accuracy of 78\%, obtained when classifying a set of approximately 42,000 users whose party registration records were available, is not as high as other methods that combine more than one type of features. But this is a much harder classification task because there are few restrictions on the users in the dataset.

\begin{table*}[th!]
\caption{Statistics on the collected datasets, including the tweets, the activity extracted from them, and the relations of the users. 
%For example, we have 4 graphs of 19,423 nodes representing the interactions that our 995 politicians had with other users in Twitter.
}\vspace{-9pt}
\resizebox{.99\textwidth}{!}
{%
\begin{tabular}{l||H H r r || r r r r || r r| r r }
%  \hline
 \multirow{2}{*}{Dataset} &
 \multicolumn{4}{c|}{\textbf{ Count }} &\multicolumn{4}{c|}{ \textbf{Activity} } 
 &\multicolumn{4}{c}{ \textbf{Relation} }\\
 \cline{2-13}
   &
  Start &
  End &
  Tweets &
  Users &
  %Nodes &
  Hashtag &
  Mention &
  Retweet &
  Quote &
  \multicolumn{2}{c}{Friends} &
    \multicolumn{2}{c}{Followers}
%   Friends & %Uniq. Fri. &
%   Followers% & Uniq. Foll
  \\
  \hline 
%\textbf{\USelection}      & Twitter & 2020-10-26 & 2021-01-04  & 348,671,076 & 20,533,417  & - & - & - & - \\

\textbf{\Public} & \small{2020-10-01} & \small{2021-02-28} & 5,804,713 & 14,112 %& 244,931 & 1,054,431 & 11,150,470 & 2,016,192  & 1,010,138 
& 1,535,221 & 9,863,454 & 4,397,565 & 420,181 & 15,452,813 & (3,700,589) & 10,619,707 & (4,050,220)\\

% \textbf{\PrePurge} & Twitter & 2020-10-26 & 2020-12-31 & 2,871,050 & 20,008 %& 244,931 & 1,054,431 & 11,150,470 & 2,016,192  & 1,010,138 
% & 255,895 & 2,510,444 & 1,621,304 & 612,827\\

\textbf{\Politicians} & \small{2020-08-01} & \small{2021-01-17}  & 156,562 & 995 %& 19,423 & 1,124,621 & 2,435,055 & 134,333 & 269,104 
& 162,121 & 212,074 & 83,920 & 54,630 & 1,724,222 & (863,509) & 33,424,367 & (9,342,089)\\
%\textbf{\Parler}      & Parler  & 2020-10-25 & 2021-01-08  & 6,546,658 & 566,486  & - & - & - & -   \\
% \hline
\textbf{\Canadian} & \small{2021-07-31} & \small{2021-10-21} & 11,361,581 & 1,114,906 & - & 15,128,727 & 8,438,765 & - & - & - & - & -
\end{tabular}
 }
    \label{table:Dataset_Description}
    \vspace{-5pt}
\end{table*}

In our experiments, we examine these two models evaluated on ``hard'' data (highlighted in bold in the table), and also the approaches of \citet{preoctiuc2017beyond} and \citet{liu2022politics}, given their performance levels, the availability of the code and that they rely on different features and data types. In particular, the method in  \citet{preoctiuc2017beyond} focuses on the use of a limited set of ideologically loaded political words. Because \citet{preoctiuc2017beyond} test their model on different types of users, the general accuracy of the model varies greatly, ranging from 62.5\% to 97.2\%. The model performs best when
%predictions are based on text content, and when 
the data is limited to users who follow politicians' account (either Democrats or Republicans).  % For example, \citet{Barber2015TweetingFL} compute the ideological position of users based on who they follow on Twitter, and use an item-response model to determine the ideology of users. 
The POLITICS approach by \citet{liu2022politics} also uses a content-based model to predict users' ideologies. Their Twitter user classification accuracy is low, under 50\%, so they are not included in Table~\ref{tab:literature_survey}. But they only tested on 3-way classification (left, right, and center), which is difficult to compare with approaches tested on 2-way (most common) or other classification tasks. %\footnote{The model used by \citet{rheault2021efficient} also classifies users according to a multi-way classification. Since they are analyzing the 2019 Canadian federal election, the model classifies users according to their party affiliation (Liberal, Conservative, New Democratic Party, Green, or People's Party of Canada) based on a test-size dataset of 505 politician accounts.}
We hypothesized that this model would have strong performance on a two-way Twitter classification task (and as shown later, this is borne out in our experiments).

We explain how each of these four baselines are implemented in the supplementary material.

\section{Methodology}
\label{sec:meth}
Here, we first discuss how we sampled the data to compare different party prediction methods, and then present our own alternative approaches, which achieve on-par or better performance and can be used based on the type of signal available.
\subsection{Data Collection}
We curated 3 datasets, summarized in Table~\ref{table:Dataset_Description} and discussed in further detail below. In this table, the users are the authors of the posts, and the activity columns represent the total occurrences of their respective activity within these posts. For example, if the numbers of posts and retweets were the same, it would imply that every post was a retweet. The relations (friends and followers) were collected separately from the posts.

\subsubsection*{\textbf{Politicians Data}}
We collected all tweets, retweets and replies from 995 public and personal Twitter accounts of the United States representatives (433), senators (99), as well as vice presidential and presidential candidates (8) using Twitter's Search API.\footnote{Some Members of Congress have more than one social medial account (e.g., one personal and one official account). In this case, we collected information for all of the relevant accounts.} We call this the \Politicians dataset. This data is collected from {2020-08-01} to {2021-01-17}.
\subsubsection*{\textbf{US Public Data}}
% \subsubsection{{Mass public data}}

We first collected around 1\% of real-time tweets using Twitter's streaming API, that included one of the following US election related keywords: [JoeBiden, DonaldTrump, Biden, Trump, vote, election, 2020Elections, Elections2020, PresidentElectJoe, MAGA, BidenHarris2020, Election2020], from {2020-10-01} to {2021-02-28} . This created a dataset (not shown in table) with approximately 350 million tweets and 20 million users. Out of these, we sample 20 thousand users from those with keywords indicating their party affiliation in their profile description: [conservative, gop, republican, trump, liberal, progressive, democrat, biden]. 
To retrieve all the information for these selected users (the 1\% collection process means missing many of a user's tweets), we then used a combination of the normal and academic Twitter APIs, to retroactively %(late 2021) 
collect the full posts and activities of the 20K users during this period. Twitter does not allow access to tweets of deleted, protected, or suspended users. Thus, we were able to successfully retrieve the tweets of about 14k users.

\paragraph{Canadian Public Data}

We also collected data on discussion of the 2021 Canadian election, from {2021-07-31} to {2021-10-21}. Similar to the US, we used a list of keywords (available in supplementary material) and collected 1\% of real-time tweets containing those keywords. Unlike the US, we did not retroactively retrieve \textit{all} tweets for the users in our sample, and examine more limited data types. It provides an illustration of some challenges one might face in applications, as well as additional experiments with a multi-party context.

\paragraph{\textbf{Additional Data}}
We further collect relation and like data on the users in our US datasets, as explained below and summarized in  Table~\ref{table:Dataset_Description}.  
\paragraph{Relations}
We collected friends (accounts a given user follows) and followers for all accounts available at collection time. For \Public, followers were retrieved during May 2022 and friends between mid July and mid August 2022. For \Politicians, both friends and followers were retrieved during January 2021. Because some users have an extreme numbers of friends or especially followers that is not feasible to retrieve, we capped the total retrieved per user at 5,000 friends and followers for normal users. For politicians we capped followers at 100,000 and retrieved friends fully (maximum friends a politician has in our data: 135,389). Given multiple users can follow the same user, we also report the number of unique friends and followers in each dataset in parentheses in Table \ref{table:Dataset_Description}. 

\paragraph{Likes}
We collected tweets liked by users in our \Public dataset in April 2022. We capped retrieval at 1000 maximum per user, leading to 5,260,616 total likes.

\subsection{Data Labeling}
To prepare our datasets for evaluation of different methods as well as for training our proposed approaches, we next need to label some of our users.\footnote{We discuss here the process for US users, which was more complex. Canadian users were labeled with a similar process through the manual labeling step, and then those manual labels were used directly (further details are available in the supplementary material).}  We first identify users with ``Republican'', ``Democrat''  keywords in their profile description. For ``Republican'' we use: [conservative, gop, republican, trump]. For ``Democrat'' we use: [liberal, progressive, democrat, biden]. We consider users as \textit{potential} ``Republican'' (``Democrat'') if the description contains at least one of the Republican (Democrat) identifiers and does not include any of the Democrat (Republican) identifiers. The rest of the users remain as ``unknown''. %Note here that we combine concepts related to both the ideology and the partisanship to label Democrat and Republican users \citep{liu2022politics}.

\subsubsection*{\textbf{Manual Labeling}}
From this set, two political science graduate students who have studied US politics manually classified  1000 general public Twitter users from each party by looking at their biographical information. Their agreement was .796 Cohen Kappa. To further improve the quality, we asked a third political science student to resolve the disagreement cases. Finally, we dropped 70 cases that could not be determined (e.g., profiles that indicated someone apolitical or independent).

\subsubsection*{\textbf{Weak Labeling}}
In order to have a larger labeled dataset for training, we classify the rest of the users as Democrats and Republicans based on the full description they provide on their \textit{user profile}. 
In particular, we use our manual labels to train a \textit{profile classifier} to generate weak labels.
This created a training set with noisy labels ($\thicksim$ 96\% accurate) that is used to train the classifiers introduced in the Proposed Methods section below, which do not rely on the user profiles and are based on activity and/or relations.  

Before training this profile classifier, we first put aside approximately one third of users which have all the necessary data types to run every approach---our main test set. Then on the remainder we finetuned a RoBERTa-large \citep{liu2019roberta} language model to predict the party each user is closest to from their profile description. %
We report the results in Table~\ref{table:user_party_alignment}. 

\begin{table}[ht]%\vspace{-10pt}
\centering
\caption{Number of users with explicit party/ideological keywords in their profile description, labeled by actual/manual label where available otherwise profile label classifier (on the left). Corresponding performance of our profile label classifier based on manually labeled sample of public users and actual politician parties (on the right).}
    \label{table:user_party_alignment}
    \vspace{-5pt}
% \small
% \resizebox{\columnwidth}{!}
{%
\begin{tabular}{l|r r | r}
%   \hline
  \multirow{2}{*}{Dataset} &
 \multicolumn{2}{c|}{ Counts } & \multirow{2}{*}{Accuracy}  \\
 \cline{2-3}
   &
  Rep.  &
  Dem.  &
    \\
  \hline 
\textbf{\Politicians} & 101  & 66   & 98.8\% \\%98.0\% & 98.5\% \\
\textbf{\Public}  & 6,355 & 7,757 & 96.0\% \\%93.0\%  &  97.0\% \\
%   \hline
\end{tabular}}\vspace{-5pt}
\end{table}

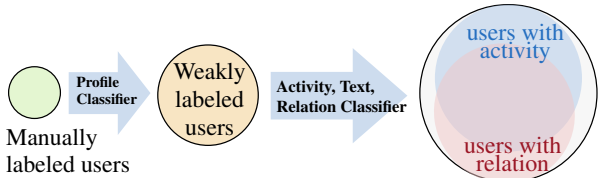
\begin{figure}
    \centering
    \resizebox{.96\linewidth}{!}{
    \tikzset{every picture/.style={line width=0.9pt}} %set default line width to 0.75pt        

\begin{tikzpicture}[x=0.75pt,y=0.75pt,yscale=-1,xscale=1]
%uncomment if require: \path (0,300); %set diagram left start at 0, and has height of 300

%Shape: Circle [id:dp9806345913877204] 
\draw  [fill={rgb, 255:red, 155; green, 155; blue, 155 }  ,fill opacity=0.08 ] (427,130.5) .. controls (427,83.28) and (465.28,45) .. (512.5,45) .. controls (559.72,45) and (598,83.28) .. (598,130.5) .. controls (598,177.72) and (559.72,216) .. (512.5,216) .. controls (465.28,216) and (427,177.72) .. (427,130.5) -- cycle ;
%Shape: Circle [id:dp060559677859620464] 
\draw  [fill={rgb, 255:red, 184; green, 233; blue, 134 }  ,fill opacity=0.38 ] (30,130) .. controls (30,116.19) and (41.19,105) .. (55,105) .. controls (68.81,105) and (80,116.19) .. (80,130) .. controls (80,143.81) and (68.81,155) .. (55,155) .. controls (41.19,155) and (30,143.81) .. (30,130) -- cycle ;
%Shape: Circle [id:dp2280190135333794] 
\draw  [fill={rgb, 255:red, 248; green, 229; blue, 195 }  ,fill opacity=1 ] (174,132.5) .. controls (174,105.71) and (195.71,84) .. (222.5,84) .. controls (249.29,84) and (271,105.71) .. (271,132.5) .. controls (271,159.29) and (249.29,181) .. (222.5,181) .. controls (195.71,181) and (174,159.29) .. (174,132.5) -- cycle ;
%Right Arrow [id:dp7906238212200162] 
\draw  [draw opacity=0][fill={rgb, 255:red, 173; green, 198; blue, 228 }  ,fill opacity=0.61 ] (88,110.25) -- (135.4,110.25) -- (135.4,90) -- (167,130.5) -- (135.4,171) -- (135.4,150.75) -- (88,150.75) -- cycle ;
%Shape: Circle [id:dp21952409625398883] 
\draw  [draw opacity=0][fill={rgb, 255:red, 147; green, 183; blue, 226 }  ,fill opacity=0.37 ] (441.5,116) .. controls (441.5,76.79) and (473.29,45) .. (512.5,45) .. controls (551.71,45) and (583.5,76.79) .. (583.5,116) .. controls (583.5,155.21) and (551.71,187) .. (512.5,187) .. controls (473.29,187) and (441.5,155.21) .. (441.5,116) -- cycle ;
%Shape: Circle [id:dp08121017414879306] 
\draw  [draw opacity=0][fill={rgb, 255:red, 240; green, 199; blue, 199 }  ,fill opacity=0.51 ] (441.25,157.46) .. controls (436.17,120.37) and (462.2,87.01) .. (499.39,82.94) .. controls (536.58,78.87) and (570.85,105.64) .. (575.93,142.73) .. controls (581.01,179.82) and (554.98,213.18) .. (517.79,217.25) .. controls (480.59,221.31) and (446.33,194.55) .. (441.25,157.46) -- cycle ;
%Right Arrow [id:dp4448018802035322] 
\draw  [draw opacity=0][fill={rgb, 255:red, 173; green, 198; blue, 228 }  ,fill opacity=0.61 ] (283,110) -- (362.2,110) -- (362.2,87) -- (415,133) -- (362.2,179) -- (362.2,156) -- (283,156) -- cycle ;

% Text Node
\draw (26,163) node [anchor=north west][inner sep=0.75pt]  [font=\huge] [align=left] {{\fontfamily{ptm}\selectfont Manually}\\{\fontfamily{ptm}\selectfont labeled users}};
% Text Node
\draw (94,113) node [anchor=north west][inner sep=0.75pt]  [font=\large] [align=left] {{\fontfamily{ptm}\selectfont \textbf{Profile}}\\{\fontfamily{ptm}\selectfont \textbf{Classifier}}};
% Text Node
\draw (287,117) node [anchor=north west][inner sep=0.75pt]  [font=\Large] [align=left] {\textbf{{\fontfamily{ptm}\selectfont Activity, Text, }}\\\textbf{{\fontfamily{ptm}\selectfont Relation Classifier}}};
% Text Node
\draw (176,100) node [anchor=north west][inner sep=0.75pt]  [font=\huge] [align=left] {\begin{minipage}[lt]{69.82pt}\setlength\topsep{0pt}
\begin{center}
{\fontfamily{ptm}\selectfont { Weakly labeled}}\\{\fontfamily{ptm}\selectfont {  users}}
\end{center}

\end{minipage}};
% Text Node
\draw (453,171) node [anchor=north west][inner sep=0.75pt]  [font=\large,color={rgb, 255:red, 155; green, 23; blue, 38 }  ,opacity=1 ] [align=left] {\begin{minipage}[lt]{93.15pt}\setlength\topsep{0pt}
\begin{center}
{\fontfamily{ptm}\selectfont\huge users with relation}
\end{center}

\end{minipage}};
% Text Node
\draw (456,62) node [anchor=north west][inner sep=0.75pt]  [font=\large,color={rgb, 255:red, 38; green, 111; blue, 192 }  ,opacity=1 ] [align=left] {\begin{minipage}[lt]{92.47pt}\setlength\topsep{0pt}
\begin{center}
{\fontfamily{ptm}\selectfont \huge users with activity}
\end{center}

\end{minipage}};

\end{tikzpicture}

% https://www.mathcha.io/editor/p3DvoC4YiMxhw5urok0xQCXNKEy0hW7jK0PCVPOEBV
    }\vspace{-10pt}
    \caption{Starting from a small set of manually labeled users, we use the profile classifier to expand to a larger set of users (our \Public dataset), and then use that to build classifiers that can classify any user with text, activity, and/or relations.}
    \label{fig:expanding_classification}
\end{figure}

We should highlight that the profile classifier, although accurate, cannot be effectively applied to users without explicit party indicators in their biographical information. This motivates methods that classify users based on other signals, e.g., who they follow or the text of their tweets. We illustrate the process of building from the small set of manually labeled users to increasingly general users in Figure~\ref{fig:expanding_classification}. We next discuss how we go from users with profile classifier labels to more general ones.

\subsection{Proposed Methods}
\label{method}\label{method:core}
Our data consists of graphs (activity and relation) and text (tweets). Here, we explain the preprocessing and models used for each data modality, namely: Graph Convolutional Networks (GCN) \cite{kipf2017semisupervised} a simple graph neural network model; Graph Attention Networks (GAT) \cite{velivckovic2018graph} a somewhat more complex one; Heterogeneous Graph Attention Networks (HAN) \cite{wang2019heterogeneous} a heterogeneous graph one; Label Propagation \cite{zhur2002learning} a non-neural graph method based on ``propagating'' label information to neighboring nodes in the graph; and RoBERTa \cite{liu2019roberta} a bidirectional transformer-based language model. These methods themselves are not new, but they have not been previously tested for this task and/or with the range of data types we examine.

\subsubsection*{\textbf{Graph Data}}

We structure the relationships between users as graphs, where the nodes are users and edges represent interactions between them. We consider two perspectives on which interactions link people together:

\begin{itemize}[leftmargin=10pt,topsep=0pt]
    \item Direct links: person $i$ mentions or retweets person $j$. This corresponds to the graph's adjacency matrix $A$, a (0,1)-square matrix, where $A_{ij}=1$ if $j$ links to $i$.
    \item Projected (indirect) links: persons $i$ and $j$ both mention the same person $k$, even though $i$ and $j$ may not mention each other directly. Formally, this corresponds to the projected adjacency matrix $A' = A^T A$ and represents co-activity.
\end{itemize}

Direct links are intuitive, however, we find empirically %(Table~\ref{tab:projection}) 
that projected links are more informative for our GCN approach. Therefore, aside from a supplementary experiment that explicitly tests this, all our \textit{GCN} models use the projected graph which represent similarity of activities between users. On the other hand, we see empirically that the direct graph performs better than the projected one for \textit{GAT}, \textit{HAN}, and \textit{Label Propagation}. So in these cases, we always report the performances on the direct graph, again excepting the experiment explicitly investigating this.

\paragraph{\textbf{GCN}} Our next step here is to construct one embedding per user and interaction type---i.e., a learned vector representing each user's realized interactions of that type (e.g., retweet, followership). We use the titular GCN \cite{kipf2017semisupervised}, which is a simple yet powerful graph representation learning model. We generally use a single-layer version that is semi-supervised, with an unsupervised link prediction task and a supervised node classification task using profile classifier labels on training nodes.

This model supports using node features as well as the graph structure. We conducted preliminary experiments on using text embeddings for this, but found it did not improve performance. This parallels \cite{xiao2020timme}, which likewise found that node features did not help their GNN-based model for party prediction. We instead use a uniform random vector with dimension 100 to initialize the embeddings.

To combine embeddings from different data types, we first train separate, independent GCNs on each individual input type, producing one embedding per user per GCN. We can then mix and match these data types as desired by concatenating the embeddings of the same user from different GCNs, before passing the combined embedding to a classifier. Here we use a random forest model to make a final prediction. In this way, we compare each data type individually and different combinations of them over multiple experiments.

\paragraph{\textbf{GAT}} Our usage of GAT is similar to GCN, except we only use node classification (no link prediction) to train it. We follow the architecture and implementation of \citet{huang2020combining}.

\paragraph{\textbf{HAN}} Our implementation of HAN is similar to GAT, but here we combine multiple graph types into a heterogeneous graph. Specifically, we combine the retweet, mention, quote, and hashtag graphs together into a single graph with those four edge types. We use the Deep Graph Library (DGL) implementation of HAN\footnote{https://github.com/dmlc/dgl/tree/master/examples/pytorch/han} which in turn is based on the original from \citet{wang2019heterogeneous}. We match the hyperparameters to our GAT models.

\paragraph{\textbf{Label Propagation}} This approach \cite{zhur2002learning} ``propagates'' labels between connected nodes (users). It is a more classical, non-neural method, and rather than an embedding, it directly produces a prediction. %It has two parameters: the number of iterations of the propagation process, and the rate $\alpha$ at which new label information replaces the old one.
We use the semi-supervised version of this algorithm, i.e., seeding train nodes with profile classifier labels. 
Unlike the preceding approaches, because there are no embeddings, it is not straightforward for Label Propagation to incorporate multiple data types. However, if the data type is chosen carefully, this simple method achieves excellent performance with very high computational efficiency as shown in Tables \ref{tab:main_experiment} and \ref{tab:compute_time}.

\subsubsection*{\textbf{Text Data}} Here, we use RoBERTa \cite{liu2019roberta}, a transformer-based \cite{vaswani2017attention} language model built on the foundation on BERT \cite{DBLP:journals/corr/abs-1810-04805}, with changes to the pretraining process to improve performance. %It uses a transformer \cite{vaswani2017attention} architecture and is pretrained on 160GB of text. 
We examine two versions, RoBERTa-base (125M parameters) and RoBERTa-large (355M parameters), and consider both an untuned version using out-of-the-box embeddings classified by a random forest, and tuning them on training set users to classify tweets according to the tweet author's label, and making a final author (user) level prediction by majority vote. 

\subsubsection*{\textbf{Further Information}} Additional details on how these approaches were implemented are available in the supplementary material under ``Supplementary Implementation Information.''
\section{Experiments}
\label{sec:exp}

We begin by illustrating the necessity of accurate party predictions through a simple experiment to examine the impact on a downstream task. Next, we move to detailed comparison of the different methods and signals for the party prediction task.

\subsection{The Impact of Party Prediction} 
% To illustrate the necessity of accurate party predictions, we do a simple experiment to examine the impact on a downstream task. In particular, 
We consider measuring political conflict, with a simple approach from the literature \cite{conover2011predicting, waugh2011party}: network modularity, corresponding to the intuition that better separated clusters reflects more \textit{political polarization}.  %The downstream task is polarization measurement. For this we take a simple approach from the literature , using the network modularity (corresponding to the intuition that better separated clusters reflects more polarization). The results are shown in Figure~\ref{fig:mod_vs_acc}.
 \setlength{\columnsep}{7pt}%
   \begin{wrapfigure}{r}{0.5\linewidth}\hspace{-15pt}\vspace{-15pt}
\begin{center}     
    \resizebox{1\linewidth}{!}{
    \includegraphics{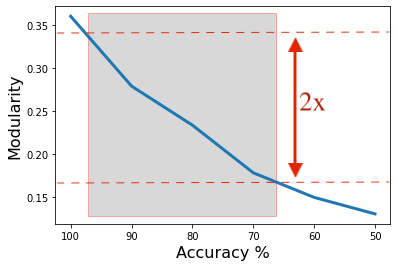}
    }\vspace{-10pt}
    \caption{\small Imperfect accuracy has a big impact on polarization measurement, i.e., the downstream task. The box represents the range of accuracies reported in the literature (Table~\ref{tab:literature_survey}).}
    \label{fig:mod_vs_acc}\vspace{-15pt}
    \end{center}
\end{wrapfigure}
We examine the measurement at different accuracy levels by taking our retweet graph data, restricted to users with manual labels, and simulating different accuracy levels by randomly swapping a percentage of the labels.
The box here represents the range of reported accuracies in the literature from Table~\ref{tab:literature_survey}. We see that the accuracy has a large effect, with the lower end of reported accuracies producing roughly half the modularity of the upper end. Consequently, low or unknown accuracy can lead to seriously misleading measurements and conclusions. This is also a simple, nearly direct measurement---the effects on more complex approaches and downstream tasks could be even more problematic.
With this motivation in mind, we next evaluate different methods to get the best predictive performance.\footnote{Because our data is relatively balanced and errors on each class matter equally in most applications, we continue to report accuracy as the performance metric. We also examined two versions of F1 (macro and weighted averaging) and found they yield similar results and conclusions.}

 \subsection{Comparing Accuracy of Methods} To compare different approaches equitably, we need a set of users that are active in all the particular ways that the different methods require. For instance, to compare \citet{barbera2015birds} (which is based  on follow relations) with \citet{preoctiuc2017beyond} (which is based on tweet text), we need users that both follow others and post tweets. We therefore require that all users in the test set have all the interaction types needed for every method we consider, to hold the test data constant across every approach.
%
% In Table~\ref{tab:main_experiment} we compare different versions of our approaches with the four benchmarks from the literature. The key to a high fidelity comparison here is holding the test data constant across every approach. 
This requires filtering for users which have all 7 interaction types considered; it also requires that they follow a political elite user (e.g., politician) to apply the Tweetscores implementation of \citet{barbera2015birds} (we relax these conditions in subsequent experiments to include more cases in our analysis). In total 655 of our manually labeled users meet this criteria. We repeat our experiments 10 times with a random sample of 40\% of these users as the test set (training each time on the fully separate profile-classifier-labeled users), and report mean and standard deviation. From the results reported in Table~\ref{tab:main_experiment}, we draw three main conclusions:

% To compare different approaches equitably, we need a set of users that are active in all the particular ways that the different methods require. For instance, to compare \cite{barbera2015birds} (which is based  on follow relations) with \cite{preoctiuc2017beyond} (which is based on tweet text), we need users that both follow others and post tweets. We therefore require that all users in the test set have all the interaction types needed for every method we consider. This filter is not applied to the training data of the corresponding methods.

\begin{table}[h]
\caption{Classification accuracy of different methods. Rank is shaded from blue (best) to orange (worst).}\vspace{-10pt}
\label{tab:main_experiment}
\resizebox{\linewidth}{!}{%
\begin{tabular}{llr}
\textbf{Type} & \textbf{Method} & \textbf{Accuracy}\\ 
\hline % this section literature methods
\multirow{6}{*}{\textbf{SOTA}} 
& %Barber\'{a} 
\citet{barbera2015birds}%} 
& 96.4 $\pm$ 0.6 \\
&  \citet{liu2022politics}: POLITICS-Untuned & 94.7 $\pm$ 0.9 \\
&  \citet{liu2022politics}: POLITICS-Finetuned & 97.8 $\pm$ 0.9
\\
& %Preo\c{t}iuc-Pietro et al. 
\citet{preoctiuc2017beyond} & 92.4 $\pm$ 0.9\\
&  \citet{xiao2020timme}: TIMME & OOM \\
\hline\hline
\multirow{2}{*}{\textbf{Text}} 
& RoBERTa-base Untuned & 86.7 $\pm$ 1.0\\
& RoBERTa-base Finetuned & 97.5 $\pm$ 1.0\\
& RoBERTa-large Finetuned & 97.6 $\pm$ 0.8 \\
\hline
\multirow{12}{*}{\textbf{Activity}} 
& Label Prop. Retweet & 97.2 $\pm$ 0.5 \\
& Label Prop. Mention & 91.0 $\pm$ 0.9 \\
& Label Prop. Quote & 95.7 $\pm$ 0.8 \\
& Label Prop. Hashtag & 92.6 $\pm$ 0.7 \\
& Label Prop. Like & 95.7 $\pm$ 0.7 \\
& GCN Retweet (RT) & 96.2 $\pm$ 1.2 \\
& GCN Mention (@) & 85.6 $\pm$ 1.7 \\
& GCN Quote (QT) & 89.4 $\pm$ 1.1 \\
& GCN Hashtag (\#) & 88.1 $\pm$ 1.8 \\
& GCN Like & 89.0 $\pm$ 1.6 \\
& GCN RT+QT & 96.3 $\pm$ 0.7 \\
& GCN All Activity (RT+QT+@+\#) & 96.3 $\pm$ 0.7 \\
& GAT Retweet (RT) & 96.9 $\pm$ 0.4 \\
& GAT Mention (@) & 88.1 $\pm$ 1.3 \\
& GAT Quote (QT) & 93.5 $\pm$ 1.0 \\
& GAT Hashtag (\#) & 91.4 $\pm$ 1.3 \\
& HAN All Activity (RT+QT+@+\#) & 90.9 $\pm$ 8.4 \\
\hline
\multirow{5}{*}{\textbf{Relation}} 
& Label Prop. Friend & 96.5 $\pm$ 0.6 \\
& Label Prop. Follow & 96.1 $\pm$ 0.8 \\
& GCN Friend & 96.5 $\pm$ 0.7 \\
& GCN Follow & 92.5 $\pm$ 1.2 \\
& GCN All Relations (Friend+Follow) & 95.2 $\pm$ 1.1 \\
& GAT Friend & 96.7 $\pm$ 0.4 \\
& GAT Follow & 95.4 $\pm$ 0.8  \\
\hline
\multirow{6}{*}{\textbf{Combined}} 
& GCN All & 96.7 $\pm$ 0.7 \\
& GCN All but Like  & 96.7 $\pm$ 0.7 \\
& GCN All but Like and Follow  & 96.8 $\pm$ 0.8 \\
& GCN RT+QT+Friend & 96.9 $\pm$ 0.8 \\
& GCN RT+Friend & 97.1 $\pm$ 0.7 \\
& GCN QT+Friend & 96.8 $\pm$ 0.8 \\
& GAT All but Like & 95.7 $\pm$ 0.8 \\
\hline
\end{tabular}%
\begin{tabular}{R}
    Rank  \\
    \hline
     15  \\
     24  \\
     1  \\
     28  \\
     - \\
     \hline     \hline
     35  \\
     3  \\
     2  \\
     \hline
     4  \\
     30  \\
     20  \\
     26  \\
     20  \\
     18  \\
     36  \\
     32  \\
     34  \\
     33  \\
     16 \\
     16 \\
     6 \\
     34 \\
     25 \\
     29 \\
     31 \\
     \hline
     13  \\
     19  \\
     13  \\
     27  \\
     23  \\
     10 \\
     22 \\
     \hline
     10  \\
     10  \\
     8  \\
     6  \\
     5 \\
     8 \\
     20 \\
     \hline
\end{tabular}
}\vspace{-10pt}
\end{table}

\begin{enumerate}[leftmargin=10pt,topsep=0pt,wide=0pt]
    \item \textbf{Testing different approaches on the same users, like here, is necessary for clear comparisons.} For example, if judging only by the number reported in their original paper, \citet{barbera2015birds} would perform over 15 percentage points worse than in our experiments. Depending on which result is being considered, none of which are close to the findings reported here, \citet{preoctiuc2017beyond} would perform anywhere from slightly better to over 30 points worse. The approach with one of the worst results for Twitter party prediction reported in the literature, POLITICS \cite{liu2022politics}, turns out to have the best performance of all.
    \item \textbf{Our relatively simple approaches deliver state-of-the-art performance.} 9 out of the top 10 approaches are ones we propose for the task and inputs here, with only a small gap between them.
    \item \textbf{One can achieve strong performance with many different types of data.} There are methods from every category within less than 1.5\% of the best method, meaning that one can get a strong prediction regardless of the category of data one has access to. Conversely, if deciding what data to collect, one can choose an efficient, scalable option (discussed in detail in subsequent experiments).
\end{enumerate}

Besides those main conclusions, we note that TIMME produced out of memory errors on this scale of data. Additional discussion of our implementation, which uses the original study's code, is available in the supplement. We also note that HAN has a high standard deviation because one of the 10 runs gave very poor performance (66.0\% accuracy). However, even with this run excluded, the average would be 93.6 $\pm$ 1.5, which would rank 25th. Finally, we note that although POLITICS Untuned is much better than RoBERTa-base Untuned, the gap after fine-tuning is slight. These are similarly-sized models---in fact the former is based on the latter. It seems most of the benefit of the special training used to produce POLITICS is erased when finetuned to this task, which is troublesome since finetuning seems necessary to get the best performance.

 \subsection{Comparing Cost of Methods}
 Here, we report the  cost of methods in terms of computational time of the models as well the time required to retrieve the data they need. 
 \subsubsection*{\textbf{Computation Time}}
Table~\ref{tab:compute_time} shows runtimes (on 1 RTX8000 GPU) for several types of approach. %\footnote{All here trained on 1 RTX8000 GPU}
We see that label propagation is multiple orders of magnitude faster than the GNN approaches. Considering label propagation also achieves very good accuracy in all our experiments, this is a very strong approach.

\begin{table}[]
\caption{Runtime for 1 run of several approaches.}
\label{tab:compute_time}\vspace{-9pt}
\resizebox{\linewidth}{!}
{%
\begin{tabular}{l|rrr}
 & \multicolumn{3}{c}{\textbf{Runtime (s)}} \\ 
\textbf{Method} & Retweet & Friend & Text \\ 
\hline
\textbf{Label Propagation} & 1 & 2& - \\
\textbf{GCN, Direct Graph} &  483 & 1,871 & - \\
\textbf{GCN, Projected Graph} &  2,682 & 5,799 & - \\
\textbf{GAT, Direct Graph} & 754 & 9,195 & -\\
\textbf{GAT, Projected Graph} & 2,374 & 4,276 & - \\
\hline
\textbf{POLITICS \cite{liu2022politics} Finetuned} & - & - & 24,370 \\
\textbf{RoBERTa-base} & - & - & 23,479 \\
\textbf{RoBERTa-large} & - & - & 55,701 \\
% \hline
\end{tabular}\vspace{-15pt}
}
\end{table}

We also note that the direct graph is generally faster than the projected graph, except for the friend graph with GAT. The number of nodes of the projected graph is fixed at the number of users of interest; only the edge density changes, and that cannot go beyond the complete graph. The direct graph though can scale arbitrarily in both number of nodes and edges. Thus at moderate graph sizes it seems the density of the projected graph requires more time to process than having more nodes with the direct graph. But with the larger friend graph (direct version has over 1.1 million nodes), the crossover point is reached for GAT and the projected version becomes faster.
Finally, we saw previously that the text-based approaches (RoBERTa and POLITICS) had strong performance, but they require much longer to train and finetune than any other option. So their applicability may be situational. Without that finetuning, their performance is very weak.

\subsubsection*{\textbf{Data Retrieval Time}} Another critical factor for efficiency is how long it takes to retrieve the data. The limiting factor here is generally the Twitter API, which imposes caps per 15 minutes. The differences are summarized in Table~\ref{tab:data_retrieval}.\footnote{Future changes to the Twitter API since this study was conducted may affect the exact numbers here. We advise future researchers to use the analysis methodology here as a guide that can be adapted to their individual data retrieval constraints.}
We see that Likes\footnote{
https://developer.twitter.com/en/docs/twitter-api/tweets/likes/api-reference/get-users-id-liked\_tweets} have a very harsh limit. Their predictive performance is also ordinary, therefore this data type is not advised unless one is already retrieving them for another purpose. Relations\footnote{https://developer.twitter.com/en/docs/twitter-api/v1/accounts-and-users/follow-search-get-users/api-reference/get-followers-ids} (i.e. friends and followers) can be retrieved in much greater number per request, 5000, leading to a ten times larger total than Likes. They do have good performance in our experiments. However, each request can only retrieve a single user. So if one has a large number of users with few relations each, the small 15 requests per 15 minutes may be a binding factor. We show this in practice in the following experiment. Finally, the limits for tweets\footnote{https://developer.twitter.com/en/docs/twitter-api/v1/tweets/timelines/api-reference/get-statuses-user\_timeline}---and therefore activity contained within tweets (i.e. retweets, mentions, quotes, and hashtags)---are the most generous. Even more can be retrieved if using the general streaming API instead of retrieving tweets for particular users. Since we also have strong methods for this type of data (e.g. label propagation on retweets), it may often be the optimal strategy.

\begin{table}[h!]
\caption{Retrieval limits for different data types}
\label{tab:data_retrieval}\vspace{-9pt}
\resizebox{\linewidth}{!}
{%
\begin{tabular}{l|rrr}
 \textbf{Data Type} & \# per Request & Requests per 15m & Total per 15m \\ 
\hline
\textbf{Tweets} & 200 & 900 & 180,000 \\
\textbf{Likes} & 100 & 75 & 7,500 \\
\textbf{Relations} & 5,000 & 15 & 75,000 \\
% \hline
\end{tabular}
}\vspace{-5pt}
\end{table}

\subsection{Comparing Coverage of Signals}
Another critical question is how many users each approach can be applied to. 
We compare the coverage of different methods in Table \ref{tab:no_allrelations_filter}, which shows the percentage of users that have the data needed to run the method.
Please note that here the test sets vary between approaches, but are always strictly larger than in experiments of Table \ref{tab:main_experiment} where we selected only users that have the data needed to apply all approaches (to keep the test set constant there). % since it include users with less diverse interactions. 
Therefore, this is a more general and harder task.
Further, we report the average users for which the needed data could be retrieved per 15 minute Twitter API cycle. Unlike the theoretical numbers in the preceding table, these numbers are calculated from the real data. Some additional notes on this and other implementation aspects are provided in the supplementary materials. 

Unsurprisingly, performance almost always decreases compared with exclusively considering users with all interaction types simultaneously. Nonetheless, we see that our methods still provide strong performance. Here RoBERTa-base is tied for top performance, and has 100\% coverage and 95\% accuracy. On the graph side, using GCN with all interactions except Like gives 100\% coverage with over 93\% accuracy, while one can get even better performance and still over 90\% coverage from several versions of label propagation, especially retweet.
Furthermore, we see that retrieving tweets (from which retweets, mentions, quotes, and hashtags can all be derived also) is far faster than retrieving friends or followers, and those two are again significantly faster than likes. Thus, using label propagation retweet or a similar method may be advisable for large-scale work. Regardless, this information can help practitioners make an informed decision.

\begin{table}[]
\caption{Accuracy on users without all relations, along with coverage (percentage of users the approach will work on), and speed (the average number of users one can retrieve per 15 minute Twitter API cycle with that data type, lower means more cost to retrieve the required data, i.e., higher data acquisition cost).}\vspace{-9pt}
\label{tab:no_allrelations_filter}
\resizebox{\linewidth}{!}{%
\begin{tabular}{llrc}
% \textbf{Relation} & \textbf{Accuracy} & \textbf{User Coverage (\%)} & \textbf{Avg. Users Retrievable per 15m}\\
\textbf{Type} & \textbf{Accuracy} & \textbf{Coverage } & \textbf{Speed }\\
\hline
% Barber\'{a} 
\citet{barbera2015birds} & 94.6 & 92.7 & 14.5 $\pm$ 0.7\\
% Preo\c{t}iuc-Pietro et al. 
\citet{preoctiuc2017beyond} & 84.7 %$\pm$ 0.0
& 100.0 & 343.2 $\pm$ 152.6\\
 \citet{liu2022politics}: POLITICS-Finetuned & 95.5 & 100.0 & 343.2 $\pm$ 152.6 \\
\hline\hline
RoBERTa-base Finetuned & 95.7 $\pm$ 0.2 & 100.0 & 343.2 $\pm$ 152.6\\
RoBERTa-large Finetuned & 95.4 $\pm$ 0.1 & 100.0 & 343.2 $\pm$ 152.6 \\
\hline
Label Prop. Retweet & 95.7 $\pm$ 0.9 & 91.2 & 343.2 $\pm$ 152.6\\
Label Prop. Mention & 87.8 $\pm$ 0.7 & 92.1 & 343.2 $\pm$ 152.6\\
Label Prop. Quote & 94.7 $\pm$ 0.5 & 80.4 & 343.2 $\pm$ 152.6\\
Label Prop. Hashtag & 90.4 $\pm$ 1.3 & 74.9 & 343.2 $\pm$ 152.6\\
Label Prop. Like & 94.1 $\pm$ 0.7 & 93.1 & 3.4 $\pm$ 5.1 \\
Label Prop. Friend & 94.8 $\pm$ 0.6 & 93.8 & 14.5 $\pm$ 0.7\\
Label Prop. Follow & 94.7 $\pm$ 0.6 & 82.2 & 14.4 $\pm$ 0.8\\
\hline
GCN All-Like  & 93.6 $\pm$ 0.9 & 100.0 & 14.4 $\pm$ 0.8\\
GCN All-Like-Follow & 94.1 $\pm$ 0.9 & 99.9 & 14.5 $\pm$ 0.7\\
GCN RT+QT+Friend & 94.6 $\pm$ 0.7 & 99.6 & 14.5 $\pm$ 0.7\\
GCN RT+Friend & 94.2 $\pm$ 1.0 & 99.6 & 14.5 $\pm$ 0.7\\
GCN QT+Friend & 93.7 $\pm$ 1.0 & 98.8 & 14.5 $\pm$ 0.7\\
GCN RT+QT & 93.9 $\pm$ 0.7 & 91.3 & 343.2 $\pm$ 152.6\\
GCN Act. (RT+QT+@+\#) & 92.7 $\pm$ 1.2 & 98.1 & 343.2 $\pm$ 152.6\\
GCN Rel. (Friend+Follow) & 92.2 $\pm$ 0.8 & 94.8 & 14.4 $\pm$ 0.8\\ 
% \hline
\end{tabular}\vspace{-15pt}
}
\end{table}

\begin{table*}[th!]
\caption{Learning from and predicting politician party affiliations. We see that politicians are slightly easier than the public. Transfer from politicians to public hurts performance slightly but is still very viable here with some approaches.}\vspace{-5pt}
\label{tab:politicians}
% \resizebox{\linewidth}{!}
{%
\begin{tabular}{ll|>{\columncolor[gray]{0.9}}ccc|c>{\columncolor[gray]{0.9}}cc}

\hline \multicolumn{2}{c|}{\textbf{Test Data}} & %\multirow{2}{*}{\textbf{Relation}} &
 \multicolumn{3}{c|}{ \cellcolor{gray!15}Public } &\multicolumn{3}{c}{ \cellcolor{gray!15} Politicians } \\ \hline
 \multicolumn{2}{c|}{\textbf{Training Data}} & Public & Politicians & Both & Public & Politicians & Both  \\
\hline
\multirow{4}{*}{\textbf{Activity}} & Label Prop. Retweet & \textbf{97.9} $\pm$ 0.9 & 95.7 $\pm$ 1.3 & 97.7 $\pm$ 1.1 & 94.5 $\pm$ 1.1 & 97.1 $\pm$ 0.9 & 96.2 $\pm$ 1.1 \\
& Label Prop. Mention & 92.0 $\pm$ 1.5 & 58.2 $\pm$ 1.5 & 90.3 $\pm$ 1.7 & 62.6 $\pm$ 2.1 & 95.8 $\pm$ 1.5 & 83.5 $\pm$ 1.3 \\
& Label Prop. Quote & 96.7 $\pm$ 1.1 & 67.8 $\pm$ 2.1 & 95.6 $\pm$ 1.1 & 77.4 $\pm$ 2.7 & 92.4 $\pm$ 1.0 & 89.1 $\pm$ 2.4 \\
& Label Prop. Hashtag & 92.6 $\pm$ 0.9 & 82.1 $\pm$ 1.5 & 92.9 $\pm$ 0.9 & 74.3 $\pm$ 3.1 & 93.5 $\pm$ 2.3 & 92.0 $\pm$ 2.5 \\
\hline
\multirow{2}{*}{\textbf{Relation}} & Label Prop. Friend & 97.5 $\pm$ 1.1 & 96.5 $\pm$ 1.0 & 97.5 $\pm$ 1.1 & 97.3 $\pm$ 0.9 & \textbf{99.7} $\pm$ 0.3 & 98.9 $\pm$ 0.5 \\
& Label Prop. Follow & 96.6 $\pm$ 1.1 & 97.4 $\pm$ 0.7 & 96.9 $\pm$ 1.0 & 98.7 $\pm$ 0.6 & 99.5 $\pm$ 0.4 & 99.4 $\pm$ 0.4 \\

\hline
\multicolumn{2}{c|}{\textbf{GCN All}} & 97.0 $\pm$ 0.7 & 91.6 $\pm$ 2.1 & 97.2 $\pm$ 0.6 & 89.8 $\pm$ 1.3 & 97.3 $\pm$ 1.5 & 95.7 $\pm$ 1.7 \\

\hline
\end{tabular}
}
\end{table*}

\subsection{Comparing Politicians \& Public}
% \paragraph{Politicians} 
In this experiment, we consider both learning from and predicting the party of politicians. Learning from politicians has a theoretical advantage because their parties are officially known, removing the need for any additional labeling process. Meanwhile, on the evaluation side, they provide an additional set of data with even more definitive labels than expert-labeled general public users. 

We show results in Table~\ref{tab:politicians}. Here the first three columns consider the task of predicting party affiliation for the general public, and compare training on the public itself, politicians, or both. The remaining three columns examine the same training scenarios but evaluated on the politicians themselves. Some additional implementation notes are given in the supplement.

Focusing first on the different interaction types (rows), we see that the best performance for politicians comes from the Friend relation (trained on politicians). This suggests politicians are especially consistent in their friends---i.e., who they follow. Considering the current strong partisan divisions in the US \citep{iyengar2019origins,mccarty2016polarized}, this agrees with the intuition that politicians may avoid following people from the opposite party in order to prevent an appearance of mixed loyalties. Meanwhile, again when trained on politicians, quote performs the worst. This could be due to politicians frequently using quotes not only to support fellow members of their party but also to rebut opponents, making this relation more challenging to learn from in this context.

Considering next the differences between datasets (columns) and results overall, performance on \Politicians is not uniformly better than on \Public, at least if the training data includes politicians (fifth column, highlighted). There is also a small but consistent improvement to predicting the public purely from adding politicians to the graph, without using their labels (first column, highlighted, compared with Table~\ref{tab:main_experiment}). Adding their labels in fact tends to hurt public prediction (third column). This may be because politicians are very well connected in the graphs, and some politicians may be targets of opposition discussion to an extent that it helps to allow the opposition label to more easily propagate through them. 

We also see that training on one user type and predicting another (second and fourth columns) can still yield satisfactory results, even though it's more challenging than training and predicting on the same type of user (first and fifth columns, highlighted). For example, one can train on politicians and get over 95\% accuracy on the general public using retweets, friends, or follows. This partially matches prior literature that there is some performance lost transferring from one type of political user to another \cite{cohen2013classifying}, but the extent of the loss is much smaller. Thus, future data and methods along these lines may enable practitioners to learn strictly from the accounts of politicians, without needing labels on the general public that are much more difficult to obtain.

\subsection{Classifying Canadian Users}

For our final experiment, we turn to Canadian politics. This illustrates challenges one might face in applications. First, the data types we had available are limited; for instance, we do not have friend or follower data. Second, we do not have all each users' tweets, only the ones collected in the 1\% real-time keyword-based sample. Third, due to time and computation limits, we examine a smaller selection of methods.

In addition, unlike the US where two parties dominate, Canada has a multi-party system. Based on the labeled data available, we examine 5-way classification [parties: Green Party (GPC), New Democratic Party (NDP), Liberal Party (LPC), Conservative Party (CPC), People's Party (PPC)]. For this and the above reasons, the task is significantly more difficult. 

We report results in Table~\ref{tab:canadian-experiment}. The test set is the same regardless of input data type used, similar to Table~\ref{tab:main_experiment}. We see a large drop in performance compared to US data. Approximately 10\% of this can be explained due to the finer-grained 5-way classification task: if we group parties into relatively left (GPC, NDP, LPC) and right (CPC, PPC), performance increases by over 10\%, e.g. Label Propagation RT+@ gives 87.2\% performance. The rest is likely due to differences in the amount of data collected per user and general differences between US and Canadian politics.

We tested here a different approach to combining data types: simply adding edges in the graph for both retweet and mention interactions (denoted in the table ``RT + @''). This enables the use of approximately 25\% more labeled train users than retweet alone, and seems to improve performance. We also see that GAT performs slightly better than label propagation. This may indicate that it can deliver benefits on this more challenging data. Nonetheless, the performance of label propagation remains competitive.

\begin{table}[]
\centering
\caption{Accuracy on Canadian users. Differences in the task and available data make this problem much harder, but one can still obtain useful results.}\vspace{-9pt}
\label{tab:canadian-experiment}
%\resizebox{\linewidth}{!}{%
\begin{tabular}{llrc}
% \textbf{Relation} & \textbf{Accuracy} & \textbf{User Coverage (\%)} & \textbf{Avg. Users Retrievable per 15m}\\
\textbf{Type} & \textbf{Accuracy} \\
\hline

RoBERTa-base Untuned & 65.3  \\
Label Prop. Retweet & 72.6 \\
GAT Retweet & 73.4 \\
Label Prop. RT+@ & 76.0 \\
GAT RT+@ & 76.5  \\
% \hline
\end{tabular}\vspace{-15pt}
%}
\end{table}

\section{Impact and Ethics}
\label{sec:impact-ethics}

Political information has potential for misuse by malicious actors. In the following we discuss impacts and ethical considerations of this work and steps taken to mitigate potential harms.

Malicious actors, such as governments attempting to interfere in other countries’ elections, are already known to spread misinformation and try to sway users to particular ideologies and voting patterns \citep{eady2023exposure}. In addition, although exact mechanisms and the role of the platform are still debated, social media has been consistently linked to increasing partisan polarization \citep{nemeth2022scoping}. Consequently, there is no security through obscurity in this domain. Strong understanding and effective countermeasures are needed to prevent the spread of misinformation and divisiveness \citep{tucker2017liberation}.

As we show in our introduction and literature review, party prediction has a fundamental role in detecting partisan biases on Twitter. To understand how groups become polarized or influenced, it is necessary to accurately identify members of these groups. Accurate party prediction can help researchers and policymakers understand how and why certain groups become polarized and how information and ideas spread within and between these groups. Although this could potentially be misused to attempt to influence users, this risk is relatively low because our methods do not enable fine-grained targeting. In addition, fine-grained targeting has already been done using other unrelated tools, such as targeted advertisements through social media platforms. Therefore, the risks are low compared to the societal benefits of better party predictions offered by our method, which should in turn lead to a better and more robust understanding of how to create healthy online environments, mitigate extreme polarization, reduce the influence of bots, and limit the spread of misinformation. 

In our study, we only use publicly available data that has been collected in compliance with Twitter's terms of service. In addition, we will never release any identifiable information in order to prevent any future misuse of the user data collected in this study. Through our collection process, all users collected publicly discuss politics in both their tweets and user profiles, and consequently would not assume their political affiliations are hidden. We note that this is an equivalent or stronger constraint on users included in our study than in many other studies in the literature (see for example \citealt{barbera2015birds,Grinberg374,pathak2021method,hobbs2017voters}).

\section{Future Work}

Our experiments showed disappointing results from methods that combined many different data types, such as TIMME, HAN, and various GCN versions. This suggests that there might be a gap in scalable methods that leverage heterogeneous graphs and/or graphs plus text to deliver superior performance in this domain. Future work testing more recent GNN methods, both homogeneous ones such as GIN \cite{xupowerful}, which might better combine graph plus text, and heterogeneous ones such as \cite{hu2020heterogeneous}, which additionally combine multiple types of graphs, would be productive. Different ways to incorporate text into these GNN models, such as different ways to produce user embeddings, could also warrant testing. Finally, entirely new methods specialized for this domain that combine multiple data types, especially ones that take scaling and data availability concerns into account, could be valuable as well.

Concurrent work done by \citet{tornberg2023chatgpt} found that ChatGPT is effective in labeling the partisanship of Twitter messages from politicians. However, since this study focused exclusively on politicians, the performance of this approach  might differ sharply on the general public \cite{cohen2013classifying}. So a comparison on general public users with some of the better-performing methods in our results, such as RoBERTa and label propagation on retweets, would be informative and help understand if and when ChatGPT labeling provides good value for this task.

In our experiments we focused on partisan users, but independent and apolitical users are also key groups in many analyses. It is more challenging to obtain ground truth labels for these users, but nonetheless this is an area for future work with many potential applications. More broadly, applying uncertainty quantification methods to party prediction might help account for less partisan users and produce more fine-grained analysis.

In this work, we focused on Twitter data, as it has been a central platform for political discourse, had good data availability, and is the subject of numerous studies in the literature. With recent content policy changes at Twitter, future data availability has become more uncertain. On the one hand, changes in API access may force researchers to reconsider what data types they use for party prediction, which adds to the value of this work. At the same time, future work extending our analysis to other platforms, such as Facebook or Reddit, would also be valuable. Similarly, we focused on US politics, which is one of the most common countries studied in the literature, as well as Canadian politics as an example of a multiparty parliamentary system. The majority of the insights here should transfer to other democracies, but further testing and adaptation to their unique characteristics would be productive to enable better party prediction, and in turn downstream research, on a more global scale.

\section{Conclusion}
\label{sec:conclusion}
Although party prediction is a foundational part of many research projects, in reviewing the literature, we found that (1) it was very challenging to compare the different methods used and that (2) they are often applied without thorough validation. To solve this problem, we first provided a survey of work on this task. This survey highlights not only the reported metrics, but also the data used, which varies widely and is a critical component of the evaluation for this task. Next, we selected state-of-the-art models from this survey and tested them on a consistent dataset we collected. Our results provide both quantitative evidence of the difficulty of comparing approaches based on the literature, and the missing thorough comparison. We also contributed multiple approaches of our own, which we studied and validated through extensive experiments, yielding insights along the way that can help further research in this area. We showed these approaches are competitive with and often out-perform state-of-the-art methods, while opening up new data types and options for practitioners. We also provided practical information about their coverage and efficiency. Our label propagation approach, particularly on retweets (or retweets plus mentions), is especially promising, with both strong performance exceeding the literature and comparable to our other best-performing approaches, and a very large efficiency (and consequently, scalability) gain compared to other approaches. This is an important finding since the common belief in the literature suggests one needs the followership graph for this task, which is much more costly to collect. Our results suggest that not only can the retweets graph be a proxy for followership relations, it in fact provides a more predictive signal for users' party affiliation. Furthermore, we also find that label propagation can be used to learn from politicians (with readily available party affiliations) to make predictions on the general public, without too large of a performance drop. Therefore, we recommend label propagation as a first tool to try for studies that need party affiliation on Twitter. We hope these experiments will provide a foundation for both applying party prediction for downstream research, and development of improved methods, such as on other platforms, or in other democracies.

\section*{Acknowledgements}

This work was partially funded by the CIFAR AI Chairs Program and by the Centre for the Study of Democratic Citizenship (CSDC). The first author is supported by funding from IVADO and by the Fonds de recherche du Qu\'ebec.

%%
%% The next two lines define the bibliography style to be used, and
%% the bibliography file.
% \bibliographystyle{ACM-Reference-Format}
\bibliography{main.bib}

% \pagebreak{4}

\appendix

\newpage

\begin{table*}[!htbp]
\caption{Appendix: Survey of methods to predict ideology (with accuracy below 65\% or no mention of the accuracy) %, as explained more in the text.  
}\vspace{-5pt}
\label{tab:literature_survey_appendix}
\resizebox{.9\linewidth}{!}
{%
\begin{tabular}{l|c|cccc|lrcc}
\textbf{Methods} &
%   \rotb{Accuracy (1)} &
%   \rotb{Media outlets (2)} &
%   \rotb{\begin{tabular}[c]{@{}c@{}}Network Activities\\   (retweets and mentions) (3)\end{tabular}} &
%   \rotb{\begin{tabular}[c]{@{}c@{}}Network Relation\\ (fellowship) (4)\end{tabular}} &
%   \rotb{\begin{tabular}[c]{@{}c@{}}Content \\ (words and  hashtags) (5) \end{tabular}} &
%   \rotb{Public vs. Elite} &
%   \rotb{Test size} &
%   \rotb{Code available?} &
%   \rotb{Level of Difficulty} \\ 
  \textbf{Accuracy} &
  \textbf{Media} &
  \textbf{\begin{tabular}[c]{@{}c@{}}Activity \end{tabular}} &
  \textbf{\begin{tabular}[c]{@{}c@{}}Relation \end{tabular}} &
  \textbf{\begin{tabular}[c]{@{}c@{}}Content  \end{tabular}} &
  \textbf{Type} &
  \textbf{Size} & % (users)} &
  \textbf{Code} &
  \textbf{Difficulty} \\ 
 \hline
\citet{liu2022politics} &
  $<$50\% &
   &
   &
   &
  \X &
  
    \begin{tabular}[c]{@{}l@{}} Public \end{tabular} &
    1,079 &
    %\href{https://github.com/launchnlp/politics}
    {\X} 
    & Hard \\
    \hline
\citet{pastor2020spotting}
&
  NA &
   &
   &
   &
  \X &
  
    \begin{tabular}[c]{@{}l@{}} Public \end{tabular} &
    20,364 &
    %\href{https://github.com/CyberDataLab/botbusters-spanish-general-elections}
    {\X} 
    & Easy \\
    \hline
\citet{sinno2022political} &
 55\% &
   \X &
   &
   &
  \X& 
  Political news &
  175 &
  &
  Easy \\
 \hline
\citet{yang2022twitter} &
   NA &
   \X &
   &
   &
  \X& 
  Public &
   NA &
  %\href{https://zenodo.org/record/6547792}
  {\X}&
  Easy
   \\
   \hline
   \citet{chen2015mass}&
   NA &
    &
   &
  \X &
  \X& 
  Both &
   NA &
  &
  Hard
   \\
   \hline
      \citet{bright2016explaining}&
   NA &
    &
   \X &
   &
  \X & 
  Both &
   NA &
   &
  Medium
   \\
   \hline
      \citet{gaisbauer2021ideological}&
   NA &
    &
   \X &
   &
  \X & 
  Both &
   NA &
   &
  Medium
   \\
   \hline
      \citet{garimella2017long} &
   NA &
   \X &
   \X &
   \X &
  \X & 
  Both &
   NA &
   &
  Medium
   \\
   \hline
    \citet{kamienski2022measuring} &
   NA &
    &
   \X &
   \X &
  \X & 
  Public &
   NA &
    &
  Easy
   \\
   \hline
\end{tabular}}\vspace{-5pt}
\end{table*}
\section{Baseline Methods}
\label{sec:baseline}
% As noted in the related work section, 
We compare our approaches with 4 state-of-the-art methods from the literature.  These were selected based on performance, taking into account the difficulty level of the data they were tested on. Here we briefly summarize some of their key points and provide the details on how we implemented them:
% and 3 simple baselines. 
   \paragraph{\textbf{Barber\'{a}}} \cite{barbera2015birds,Barber2015TweetingFL} This item response model assigns scores to users based on who they follow.
   We use the Tweetscores\footnote{https://github.com/pablobarbera/twitter\_ideology} implementation, which compares a user against ``elite'' (politicians and media) users with pre-trained scores. 
   
   \setlength{\columnsep}{7pt}%
   \begin{wrapfigure}{r}{0.5\linewidth}\hspace{-15pt}\vspace{-15pt}
\begin{center}    % \resizebox{.5\linewidth}{!}{
    \includegraphics[width=1\linewidth]{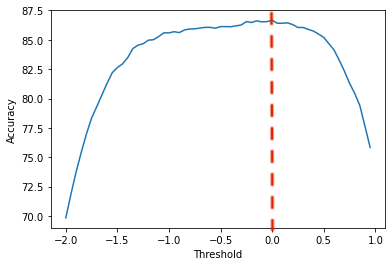}
    % }
    \end{center}\vspace{-10pt}
    \caption{\small Tuning the threshold for classification with Barber\'{a}'s approach. Accuracy is maximized at exactly 0.}
    \label{fig:barbera_threshold}\vspace{-10pt}
\end{wrapfigure}
   We classify any user with score greater (resp. less) than 0 as Republican (resp. Democrat), which both matches intuition and gives the best performance empirically. % (see Appendix~\ref{app:appendix_c}).
% \section{Appendix C: Tuning Classification Threshold for Barber\'{a}}
% \label{app:appendix_c}
In particular, we show here the results of different classification thresholds for Barber\'{a}'s model. This shows a single run in the setting of Table~\ref{tab:no_allrelations_filter} (all users available), with threshold increments of 0.05 from -2 to 1. %Anything user below the threshold is classified as Democrat and any user above as Republican. We see that the intuitive threshold of 0.0 performs best.
% \begin{itemize}
\vspace{-5pt}
    
  \paragraph{\textbf{Preo\c{t}iuc-Pietro}} \cite{preoctiuc2017beyond} This is a bag-of-words style approach with a custom, specialized vocabulary of 352 politics-related words. For each user we concatenate all their tweets into one document and calculate the counts of these words. Following the original paper, we then use this feature vector as input to a logistic regression that classifies each user. We implement the logistic regression through scikit-learn \cite{scikit-learn} with default hyperparameters.
    \vspace{-5pt}
\paragraph{ \textbf{POLITICS}} \cite{liu2022politics} This model is based on RoBERTa-base, but adds political domain adaptation through training on news articles with ideology labels. We pass the final-layer embedding of the [CLS] token through a single fully-connected one-layer classification head. With this architecture, we fine-tune the model with the same setup that we used for RoBERTa-base. We also report performance of an untuned version where we obtain user embeddings from averaging the pretraining-only tweet chunk [CLS] embeddings, and classifying users in the same way that the we classify the user embeddings from GCN.

  \vspace{-5pt}
\paragraph{\textbf{TIMME}} \cite{xiao2020timme} This approach is based on a variation of a GCN, but the architecture is adapted to learn from multiple data types simultaneously and end-to-end. This contrasts with our GCN approach which trains an independent GCN per data type, and combines them in a separate final stage. Therefore, this design hopes to learn more complex interactions between the different data types, at the cost of being much more computationally intensive and potentially more difficult to train. Unfortunately, even when using high-end AI-specialized hardware and the original authors' code,\footnote{https://github.com/PatriciaXiao/TIMME} we found the computational burden is severe. We elaborate in the following note. % section.
% \end{itemize}

% The simple baselines were selected to cover both graph-based and text-based approaches:

% \begin{itemize}
%     \item \textbf{RoBERTa-base} \cite{liu2019roberta} This language model is based on BERT \cite{DBLP:journals/corr/abs-1810-04805}, with changes to the pretraining process to improve performance. It uses a transformer \cite{vaswani2017attention} architecture and is pretrained on 160GB of text.
%     \item \textbf{RoBERTa-large} \cite{liu2019roberta} This is a larger version of the above, with 355M parameters instead of 125M.
%     \item \textbf{Label Propagation} \cite{zhur2002learning} This graph algorithm labels nodes iteratively based on the majority label of their neighbors. Compared with our GCN method, this approach is more classical and non-neural. We tested number of iterations in [1,2,3,4,5,6,8,10,15,20,25] and found performance decreases monotonically with more iterations at $\alpha$ in [0.1,0.5,0.9]. Therefore, unless otherwise noted, we use a single iteration in the experiments, which amounts to directly taking a majority vote of neighbors.
% \end{itemize}
\subsection{A Note on TIMME model}\label{sec:timmetime}
We attempted to implement TIMME using 100GB RAM and an RTX8000 GPU (which has 48GB VRAM), using the code released by the authors. However, we found both versions TIMME and TIMME-Hierarchical frequently produced out of memory errors with our scale of data. This is likely because, unlike all our proposed graph-based approaches, TIMME requires loading multiple graphs in memory simultaneously for end-to-end training. We also note that our data has over 4 times more interactions than the largest dataset used in the TIMME paper, and over 100 times more than the smallest. In addition, in some tests where we restricted the data and got it to run, it was over 100 times slower per epoch than our GCN approach (which is itself 100 times slower than our label propagation approach), and did not give good performance. Therefore, although this is arguably the most sophisticated approach and might give strong performance in some settings like the original paper, the current version is challenging to apply to larger datasets. 

\section{Distribution of Unretrievable US Public Users}

We examine the distribution of the missing users in Table~\ref{table:missing_users}.
%\footnote{We omit from the table four users whose retrieval failed for other miscellaneous reasons, e.g. Twitter API crash. We also omit 67 users whose profile label was corrupted due to a data storage error. These errors have no correlation with the dimensions examined in the table, and the users are too few to impact any conclusions drawn from it.}
We see that a strong majority are Republicans. We hypothesize that many of these users left around the January 6$^{th}$ capitol attack, either voluntarily or during the subsequent wave of suspensions. % that included the account of Donald Trump.

\begin{table}[ht]\vspace{-5pt}
\centering
\caption{Distribution of users whose data could not be retrieved retroactively. 
The majority are Republican, possibly users who left around January 2021, i.e., the wave of suspensions that included Donald Trump.} \vspace{-9pt}
% \small
% \resizebox{\columnwidth}{!}{%
\begin{tabular}{l|r r }
% \hline
  %Party
  &
  \textbf{Republican}  &
  \textbf{Democrat}  \\
  \hline 
\textbf{Suspended} & 1,824   & 439   \\
\textbf{Deleted}  & 1,857 & 543 \\
\textbf{Private} & 186 & 267 \\
%   \hline
\end{tabular}\vspace{-5pt}
% }
    \label{table:missing_users}
\end{table}

\section{Distribution of Retweets and Quotes}

In Table~\ref{table:rtqt-dist} we show the proportion of tweets that are retweets or quotes. We see that there is a similar proportion of retweets between the US and Canadian public datasets. This is intuitive since both datasets contain users from the general public and are sampled with relatively similar processes aside from the country difference. On the other hand, the US Politicians show a clear difference in behavior compared to the public -- they retweet less and quote more.

\begin{table}[ht]\vspace{-5pt}
\centering
\caption{Proportions of tweets that are retweets or quotes. US and CA public are similar, while politicians retweet less and quote more.} \vspace{-9pt}
% \small
% \resizebox{\columnwidth}{!}{%
\begin{tabular}{l|r r r }

  & Tweets & \% Retweet & \% Quote \\
  \hline 
\textbf{US Public} & 5,804,713 & 75.8 & 7.2   \\
\textbf{US Politicians}  & 156,562 & 53.6 & 34.9\\
\textbf{CA Public} & 11,361,581 & 74.3 & - \\
%   \hline
\end{tabular}\vspace{-5pt}
% }
    \label{table:rtqt-dist}
\end{table}

\section{Additional Information on Canadian Data}

\paragraph{Collection} We sampled 1\% of real-time tweets containing at least one of the following keywords:

{\scriptsize `trudeau', `freeland', `o’toole', `bernier', `blanchet', `jagmeet singh', `annamie', `debate commission', `reconciliation', `elxn44', `cdnvotes', `canvotes', `canelection', `cdnelection', `cdnpoli', `canadianpolitics', `canada', `forwardforeveryone', `readyforbetter', `securethefuture', `NDP2021', `votendp', `orangewave2021', `teamjagmeet', `UpRiSingh', `singhupswing', `singhsurge', `VotePPC', `PPC', `peoplesparty', `bernierorbust', `mcga', `saveCanada', `takebackcanada', `maxwillspeak', `LetMaxSpeak', `FirstDebate', `frenchdebate', `GovernmentJournalists', `JustinJournos', `everychildmatters', `votesplitting', `ruralcanada', `debatdeschefs', `électioncanadienne', `polican', `bloc', `jevotebloc'}

\paragraph{Labeling} From the main dataset, we sampled users with the following keywords in their profile:\\
CPC: {\scriptsize `erin o'toole', `andrew scheer', `conservative', `conservative party', `cpc', `cpc2021', `cpc2019', `conservative party of canada'} \\
GPC: {\scriptsize `annamie paul', `green party', `gpc', `gpc2019', `gpc2021', `green party of canada'} \\
LPC: {\scriptsize `justin trudeau', `liberal', `liberal party', `lpc', `lpc2021', `lpc2021', `lpc2019', `liberal party of canada'} \\
NDP: {\scriptsize `jagmeeet singh', `new democrat', `new democrats', `new democratic party', `ndp', `ndp2021', `ndp2019'} \\
PPC: {\scriptsize `maxime bernier', `people's party', `ppc', `ppc2019', `ppc2021', `people's party of canada'}

Two political science graduate students who have studied Canadian politics then manually verified or corrected the labels. This led to the counts of labeled users shown in Table~\ref{table:canadian_label_counts}

\begin{table}[ht]\vspace{-5pt}
\centering
\caption{Counts of manually-labeled Canadian users.} \vspace{-9pt}
% \small
% \resizebox{\columnwidth}{!}{%
\begin{tabular}{l|r  }
% \hline
  Party
  & Count\\
  \hline 
  
CPC &   2039\\
LPC  &  1808\\
NDP   &  704\\
PPC    & 451\\
GPC     & 299\\
%   \hline
\end{tabular}\vspace{-5pt}
% }
    \label{table:canadian_label_counts}
\end{table}

\section{Supplementary Implementation Information}
\label{app:implementation}

For each individual interaction type, we train only on the top 50\% of most active users for that relation according to raw counts. This filter is not applied to test data.

\paragraph{GCN}
 We train for 1000 epochs with the Adam optimizer \cite{kingma2014adam} with PyTorch default parameters (learning rate = 1e-3). When we use a random forest to combine different data types, we use the scikit-learn \cite{scikit-learn} implementation with default hyperparameter settings. As in all experiments, we train it using users with profile classifier labels, and report test results on a (fully separate) set of users using manual labels. 

\paragraph{GAT} We use the implementation and hyperparameters from https://github.com/xnuohz/CorrectAndSmooth-dgl corresponding to \cite{huang2020combining}. This includes hyperparameters of 2000 epochs of training, learning rate 0.002, dropout 0.75, and 3 layers. We use hidden dimension 100 matching our GNN models. We slightly adjust the last GAT layer to exactly match the others (i.e. GATConv, linear, batchnorm, activation, dropout), followed by a final linear layer, bias, and softmax. This produces practical final layer embeddings, similar those we obtain from the GCN, which we tested in Table~\ref{tab:semi_vs_unsupervised}. 

\paragraph{HAN} We use the implementation of https://github.com/dmlc/dgl/tree/master/examples/pytorch/han which is a reimplementation of the original HAN paper \cite{wang2019heterogeneous}. For the hyperparameters, we found that the ones given originally, applied to our data, resulted in collapse of the predictions to a single class. Matching our GAT hyperparameters, however, gave more effective results reported in the main experiments. Specifically, we matched learning rate, dropout, number of heads, hidden units, and number of epochs. We left the default 20 batch size, and removed weight decay for both a fairer comparison with our other models which do not use it, and because all nonzero values we tested (including the default) resulted in class collapse of the predictions.

For the metapaths, we use the basic ones corresponding to the edge types: retweet, mention, quote, and hashtag. We also tested using only retweets and mentions, and only retweets and quotes. However, these proved significantly more unstable than using all four together (resulting in more cases of class collapse), and with the exception of two runs of the ten using all four edge types resulted in equal or better performance. Of the two cases where it did not, one was when using all four resulted in class collapse, while in the other the margin was slight (in favor of retweet plus quote by less than half a percentage point).

\paragraph{Label Propagation} There are two parameters: the number of iterations of the propagation process, and the rate $\alpha$ at which new label information replaces the old one. We first tested it on projected graphs, where we found performance decreases consistently and monotonically with more iterations irrespective of $\alpha$ value (specifically, tested iterations in [1,2,3,4,5,6,8,10,15,20,25], with fully tested alpha values [0.1,0.5,0.9] and partially tested [0.03,0.97]). When using a single iteration, this method corresponds to taking the majority vote of the neighbors' labels, and changing $\alpha$ has no effect.

We later found that performance is better on the direct graph, specifically with two iterations and $\alpha = 0.5$. With 1 iteration on this graph, the performance is poor, while with more iterations performance remains constant (accounting for margin of error) or decreases. The direct graph takes much longer to run, so due to time/computation constraints we were not able to test more values of $\alpha$ in this setting. Therefore, except where stated otherwise, we report results from the best performing version, i.e., two iterations on direct graph with $\alpha = 0.5$.

\paragraph{RoBERTa} In order to provide additional context for the models, We concatenated each user's tweets in time order into chunks. Since the positional encoding scheme of these language models limits the length of input to a fixed number of tokens, we start a new chunk each time the previous chunk goes above the length limit of the model. Any tweets that don't fit into a single chunk are truncated and placed into a chunk of their own. As a result, no tweet are split between chunks. Note that we also use this same preprocessing for the POLITICS \cite{liu2022politics} baseline model described previously, which has a similar architecture.

These models are pretrained without a sequence classification task. Consequently, despite RoBERTa models having a [CLS] token like BERT and many similar language models, the representation of this token is not pretrained and thus unsuitable for direct use in downstream tasks. In some experiments we tested the model without the finetuning needed to properly learn this token. Therefore, to improve comparability, in all experiments with RoBERTa instead of [CLS] we use the mean of all individual word embeddings from the final output layer.

To further improve the prediction accuracy, we fine-tuned these models with a tweet chunk classification task. First, we label each chunk in the train set according to the profile classifier (weak) label of the corresponding user. We add a fully-connected dense layer (the ``classification head") to each model. The input size of this dense layer is equal to the size of the hidden dimensions of each transformer model, while the output size of this layer is equal to the number of label classes ($2$ in our case.).
We then finetuned each model end-to-end for $1$ epoch using the Optax \cite{deepmind2020jax} implementation of the adamw optimizer \cite{loshchilov2017decoupled} with default weight decay strength 1e-4. We set the learning rate to 1.592e-5 for RoBERTa-base, 2.673e-5 for POLITICS, and 4.269e-6 for RoBERTa-large, based on gridsearch hyperparameter tuning on a validation set of users labeled by our profile classifier. When evaluated on the profile classifier labels, this setup leads to $91.5\% \pm 0.2\%$ test accuracy on tweet chunks for RoBERTa-base, $91.6\% \pm 0.3\%$ for POLITICS and $91.9\% \pm 0.3\%$ for RoBERTa-large. Note that these accuracies for tweet chunk classification are only on weak labels, and are correlated with but not directly comparable to user classification reported in all other experiments.

To get a prediction for each user, instead of a prediction on an individual tweet chunk, we first predict a label for each of a user's chunks. Then we take the majority vote. 

\section{Additional Experiments}
%We next conduct two important ablation experiments.
\subsection{Effect of Supervision} We compare our GCN model, which does both unsupervised link prediction and supervised party prediction with training data labeled by the profile classifier, with training embeddings through a fully unsupervised GCN doing link prediction alone. We also evaluate the difference between using the supervision in the graph model itself versus sending the embeddings to a random forest (RF) for the final prediction.

\begin{table}[]
\caption{Comparing semisupervised (SS) vs. unsupervised (US) GCN. Semisupervised performs better.}\vspace{-5pt}
\label{tab:semi_vs_unsupervised}
\resizebox{\linewidth}{!}{%
\begin{tabular}{lccccccccccc}
 & Retweet & Mention & Quote & Hashtag & Friend & Follow \\%& All \\ 
\hline
\textbf{GCN SS}  & 96.7 $\pm$ 0.7  & 83.7 $\pm$ 1.8 & 90.3 $\pm$ 1.5  & 88.7 $\pm$ 1.8 & 96.8 $\pm$ 0.6 & 93.3 $\pm$ 1.1 \\%& NA \\
\textbf{GCN SS+RF}  & 96.2 $\pm$ 1.2  & 85.6 $\pm$ 1.7 & 89.4 $\pm$ 1.1  & 88.1 $\pm$ 1.8 & 96.5 $\pm$ 0.7 & 92.5 $\pm$ 1.2 \\%& 96.7 $\pm$ 0.7\\
\textbf{GCN US+RF} & 96.1 $\pm$ 0.6 & 83.1 $\pm$ 2.2 & 87.6 $\pm$ 1.2 & 87.1 $\pm$ 1.3 & 94.0 $\pm$ 1.1 & 91.9 $\pm$ 0.9 \\%& 96.1 $\pm$ 0.9 \\
\hline
\textbf{GAT SS} & 96.9 $\pm$ 0.4 & 88.1 $\pm$ 1.3 & 93.5 $\pm$ 1.0 & 91.4 $\pm$ 1.3 & 96.7 $\pm$ 0.4 & 95.4 $\pm$ 0.8 \\%& NA \\
\textbf{GAT SS+RF} & 91.9 $\pm$ 1.6 & 73.3 $\pm$ 3.5 & 81.9 $\pm$ 1.6 & 78.2 $\pm$ 3.4 & 88.3 $\pm$ 1.1 & 86.3 $\pm$ 1.2 \\%& 92.6 $\pm$ 2.5 \\
\hline
\textbf{Label Prop.}
& 97.2 $\pm$ 0.5 & 91.0 $\pm$ 0.9 & 95.7 $\pm$ 0.8 & 92.6 $\pm$ 0.7 & 96.5 $\pm$ 0.6 & 96.1 $\pm$ 0.8 \\%& NA \\
\hline
\end{tabular}\vspace{-15pt}
}
\end{table}

Results are shown in Table~\ref{tab:semi_vs_unsupervised}. First, we see that although the margin is not huge, the semisupervised GCN version performs consistently better. This indicates the supervision helps to produce more informative embeddings from the GCN. 

Second, we see that the RF, although it enables one to combine different interaction types together as in the experiments above, hurts performance slightly with GCN. With GAT, it is significantly detrimental, possibly due to architectural differences. However, even without the RF, both GCN and GAT are outperformed by label propagation in almost every case.\footnote{Consequently, in other experiments, unless otherwise noted, we report all GCN results using the semisupervised version with RF, so the results are comparable when combining data types. While for GAT, which we seldom use for combining data types, we report the version without RF.}

\subsection{Effect of Projection} In this experiment, we examine the impact on performance of using projected (indirect) graphs vs. the original (direct) ones. We show results in Table~\ref{tab:projection}.

\begin{table}[]
\caption{Comparing projected (Pro.) vs. direct (Dir.) graphs. Projection improves GCN performance. But it hurts label propagation (LP) performance, at least if the number of iterations is tuned for it.}\vspace{-5pt}
\label{tab:projection}
\resizebox{\linewidth}{!}{%
\begin{tabular}{lccccccccccc}
 & Retweet & Mention & Quote & Hashtag & Friend & Follow \\% & All \\ 
\hline
\textbf{Pro. GCN-1L}  & 96.2 $\pm$ 1.2 & 85.6 $\pm$ 1.7 & 89.4 $\pm$ 1.1 & 88.1 $\pm$ 1.8 & 96.5 $\pm$ 0.7 & 92.5 $\pm$ 1.2 \\%& 96.7 $\pm$ 0.7 \\
\textbf{Dir. GCN-1L} & 92.1  $\pm$ 0.8 & 77.1 $\pm$ 2.3 & 83.1 $\pm$ 1.7 & 79.1 $\pm$ 1.7 & 88.9 $\pm$ 1.8 & 75.1 $\pm$ 2.8 \\%& 94.4 $\pm$ 1.1 \\
\textbf{Dir. GCN-2L} & 87.4 $\pm$ 1.5 & 71.2 $\pm$ 2.3 & 76.3 $\pm$ 2.6 & 76.3 $\pm$ 1.3 & 82.4 $\pm$ 2.8 & 69.3 $\pm$ 2.7 \\%& 88.0 $\pm$ 1.8 \\
\hline
\textbf{Pro. GAT} & 
96.6 $\pm$ 0.6 & 66.9 $\pm$ 1.3 & 92.2 $\pm$ 1.3 & 90.2 $\pm$ 1.6 & 94.5 $\pm$ 0.9 & 96.1 $\pm$ 0.9 \\%& NA \\
\textbf{Dir. GAT} & 
96.9 $\pm$ 0.4 & 88.1 $\pm$ 1.3 & 93.5 $\pm$ 1.0 & 91.4 $\pm$ 1.3 & 96.7 $\pm$ 0.4 & 95.4 $\pm$ 0.8 \\%& NA\\
\hline
\textbf{Pro. LP-1I} & 96.6 $\pm$ 0.6 & 66.5 $\pm$ 1.9 & 83.4 $\pm$ 0.9 & 86.3 $\pm$ 1.1 & 70.1 $\pm$ 1.6 & 96.6 $\pm$ 0.8 \\%& NA\\
\textbf{Dir. LP-1I} & 75.4 $\pm$ 1.9 & 68.1 $\pm$ 1.8 & 68.7 $\pm$ 1.9 & 66.3 $\pm$ 1.9 & 88.0 $\pm$ 1.4 & 87.3 $\pm$ 1.5 \\%& NA \\
\textbf{Dir. LP-2I} & 97.2 $\pm$ 0.5 & 91.0 $\pm$ 0.9 & 95.7 $\pm$ 0.8 & 92.6 $\pm$ 0.7 & 96.5 $\pm$ 0.6 & 96.1 $\pm$ 0.8 \\%& NA \\
\hline
\end{tabular}\vspace{-15pt}
}
\end{table}

We see that GCN performance degrades when not projecting the graph. Adding another layer to the GCN, which in principle might help it use information from two-hop neighbors in a similar way to projection, turns out to be further detrimental. On the other hand, for GAT and label propagation, the direct graph generally performs better. The exception is the follow relation, which in all cases is better with the projected graph, which might reflect underlying user behavior differences between the interaction types. For label propagation in particular, the best version we found is two iterations on the direct graph. The best projected graph version is one iteration, but it is clearly worse overall, while one iteration on the direct graph is worse still.

Overall, along with the earlier results on runtime, the choice of direct vs. projected graph can have a significant impact. It is not as commonly tested as simple hyperparameters, and may be worth examining in more contexts.

\section{Users without all interaction types experiment: additional notes}
\label{sec:no_allrelations_notes}

Average users retrievable is calculated by first taking the per user counts of the necessary data type in our dataset. From there and the numbers in Table~\ref{tab:data_retrieval}, we calculate how many API requests are needed to retrieve that user's data, and finally how many users can be retrieved in every 15 minutes. Retrieval from different API endpoints can be run in parallel, so when an approach uses a combination of data types, we report on the one that takes the longest to retrieve.

The standard deviation reported for our approaches here is the result of re-running the model itself 10 times, as in other experiments, but here because we examine all possible test users the test set does not change within these 10 runs. Existing state-of-the-art models were run once. The standard deviation for users retrievable is not from multiple runs; rather, it is the standard deviation of our sample in its estimation of the average users retrievable.

When evaluating combinations of interaction types we use the GCN embedding where available; otherwise we treat the embedding as all zeros. So for example, if we are considering Retweet plus Quote (RT+QT in Table~\ref{tab:no_allrelations_filter}), then if both types of activity are available we use both. While if only one is available we concatenate that one with a vector of zeroes for the missing one, which tells the final classification model that quote is not available.

\section{Politicians experiment: additional notes}
\label{sec:politicians_notes}

In this experiment we use the unsupervised GCN version of our model. This lets us test how training the final classification on \Politicians alone will translate to performance on \Public, with the GCN part of our model (and thus the embeddings) held constant.

When testing on \Public using \Politicians, we use all of them, while when testing on \Politicians, we do a 75-25 random split. In all cases the setting corresponds to Table~\ref{tab:main_experiment}, i.e. users with all interaction types, again to keep the test set consistent between the different approaches.

% \section{Tweets vs. Tweet Chunks} 

% In Table~\ref{tab:tweets_vs_chunks}, we compare the technique of concatenating tweets to provide additional context, vs. embedding a single tweet at a time. We see that the former gives more accurate embeddings. 

% Note that this experiment was run prior to finetuning, using the final random forest part of the GCN model to make the prediction with the untuned RoBERTa embeddings as input. Therefore, the language model part of this process is only a single run (there is only one pretrained model available), and the variance reported is only from the random forest part. Thus this experiment also shows the improvement in the model due to finetuning: accuracy increases from 83.6\% here to 89.2\% in Table~\ref{tab:main_experiment}. 

% \begin{table}[]
% \caption{Comparing tweets vs. tweet chunks. The latter technique improves performance.}
% \label{tab:tweets_vs_chunks}
% \resizebox{\linewidth}{!}{%
% \begin{tabular}{lcc}
%  & Tweet & Tweet Chunk \\ 
% \hline
% %\textbf{RoBERTa-base - untuned} & 81.9 $\pm$ 2.3 & 83.6 $\pm$ 1.5 \\
% \textbf{RoBERTa-base - untuned} & - $\pm$ - & - $\pm$ - \\
% \hline

% \end{tabular}
% }
% \end{table}

\section{Computational Resources and Libraries}

%Code and data are available at [INSERT LINK]. 
Most experiments were done using RTX8000 GPUs. A number of text-based experiments were also run on v3-8 and v4-8 TPU VMs. Graph-based models were implemented using DGL \cite{wang2019dgl}, and language models using HuggingFace \cite{Wolf2019HuggingFacesTS} in JAX \cite{jax2018github}. To run all models of the main comparison experiment (Table~\ref{tab:main_experiment}), %excluding TIMME (which is very slow; please see details in experiments section), 
the roughly estimated time using 10 RTX8000 is one to two weeks.

\section{Additional Literature Summary}
\label{app:appendix_a}

In Table~\ref{tab:literature_survey_appendix} we provide information on papers with lower or not reported performance to predict ideology, below the threshold for inclusion in Table~\ref{tab:literature_survey}. In contrast with Table 1, this survey includes only papers for which the accuracy was less than 65\% or unknown.

\end{document}